\documentstyle[12pt,amssymbols]{article}
\newcommand{\be}{\begin{equation}}
\newcommand{\ee}{\end{equation}}
\newcommand{\bea}{\begin{eqnarray}}
\newcommand{\eea}{\end{eqnarray}}
\newcommand{\brr}{\begin{array}}
\newcommand{\err}{\end{array}}
\newcommand{\bc}{\begin{center}}
\newcommand{\ec}{\end{center}}
\newcommand{\nn}{\nonumber}
\newcommand{\sss}{\scriptscriptstyle}

\newcommand{\epuno}{\epsilon}
\newcommand{\epdue}{\epsilon^{ 2}}
\newcommand{\gammazeros}{\hat{\gamma}^{(0)T}_{ s}}
\newcommand{\gammazeroe}{\hat{\gamma}^{(0)T}_{ e}}
\newcommand{\gammaones}{\hat{\gamma}^{(1)T}_{ s}}
\newcommand{\gammaonee}{\hat{\gamma}^{(1)T}_{ e}}

\newcommand{\mt}{m_{ t}}
\newcommand{\Mw}{M_{ W}}

\newcommand{\V}{\hat{ V}}
\newcommand{\J}{\hat{J}}
\newcommand{\Ke}{\hat{ K}}
\newcommand{\U}{\hat{U}}

\newcommand{\W}{\hat{     W}}
\newcommand{\Z}{\hat{     Z}}

\newcommand{\PP}{\hat{ P}}

\newcommand{\xt}{x_{ t}}

\newcommand{\Gfive}{\gamma_{ 5}}
\newcommand{\Gmuup}{\gamma^{\mu}}
\newcommand{\Gmudw}{\gamma_{\mu}}
\newcommand{\Gnuup}{\gamma^{\nu}}
\newcommand{\Gnudw}{\gamma_{\nu}}
\newcommand{\Group}{\gamma^{\rho}}
\newcommand{\Grodw}{\gamma_{\rho}}

\newcommand{\Gmulup}{\gamma^{\mu}_{ L}}

\newcommand{\LL}{\gamma^{\mu}_{ L}\otimes\gamma_{\mu L}}
\newcommand{\LR}{\gamma^{\mu}_{ L}\otimes\gamma_{\mu R}}

\newcommand{\RRn}{\gamma^{\nu}_{ R}\otimes\gamma_{\nu R}}
\newcommand{\SP}{(1\!-\!\Gfive)\otimes(1\!+\!\Gfive)}
\newcommand{\LLR}{\gamma^{\mu}_{ L}\otimes\gamma_{\mu L,R}}

\newcommand{\GLup}{\Gamma^{\mu}_{ L}}

\newcommand{\GRup}{\Gamma^{\mu}_{ R}}

\newcommand{\GLRdw}{\Gamma_{\mu L,R}}
\newcommand{\GLLdw}{\Gamma_{\mu L}}
\newcommand{\alphas}{\alpha_{ s}}
\newcommand{\alphae}{\alpha_{ e}}
\newcommand{\iii}{\Bbb I}

\newcommand{\LV}{\gamma^{\mu}_{ L}\otimes\gamma_{\mu}}
\newcommand{\LA}{\gamma^{\mu}_{ L}\otimes\gamma_{\mu}\Gfive}
\def\V{\rule{0pt}{2ex}}
\def\tsp{\rule{0pt}{2.1ex}}
\def\tsq{\rule{0pt}{2.5ex}}

\begin{document}
\setcounter{page}{1}
\begin{flushright}
LPTENS 93/11 \\
ROME prep. 92/913 \\
ULB-TH 93/03\\
April 1993\\
\end{flushright}
\centerline{\bf{The $\Delta S=1$  Effective Hamiltonian Including}}
\centerline{\bf{Next-to-Leading Order  QCD and QED Corrections}}
\vskip 1cm
\centerline{\bf{ M. Ciuchini$^{a,b}$,
E. Franco$^b$, G. Martinelli$^{b,c}$
and L. Reina$^d$ }}
\centerline{$^a$ INFN, Sezione Sanit\`a, V.le Regina Elena 299,
00161 Roma, Italy. }
\centerline{$^b$ Dip. di Fisica,
Universit\`a degli Studi di Roma ``La Sapienza" and}
\centerline{INFN, Sezione di Roma, P.le A. Moro 2, 00185 Roma, Italy. }
\centerline{$^c$ Laboratoire de Physique Th\'eorique de }
\centerline{l'\'Ecole Normale Sup\'erieure\footnote
{Unit\'e Propre du Centre National de
la Recherche Scientifique,
 associ\'ee \`a  l'\'Ecole Normale Sup\'erieure  et \`a
l'Universit\'e de Paris-Sud.}}
\centerline{24 rue Lhomond, 75231 Paris CEDEX 05, France.}
\centerline{$^d$ Service de Physique Th\'eorique,
Universit\'e Libre de Bruxelles,}
\centerline{Boulevard du Triomphe, CP 225 B-1050 Brussels, Belgium.}
\begin{abstract}
In this paper we present a calculation of the $\Delta S=1$
 effective weak Hamiltonian
including next-to-leading order QCD and QED  corrections.
At a scale $\mu$ of the order of few GeV, the Wilson
coefficients of the operators are given
  in terms of the renormalization group evolution
matrix  and of the coefficients computed at a large scale $\sim M_W$.
The expression of the evolution matrix is derived from
the two-loop anomalous dimension matrix which governs
the mixing of the relevant current-current and penguin operators,
renormalized in some given  regularization scheme.
We have computed the anomalous dimension matrix up to and
including order $\alphas^2$ and $\alphae \alphas$ in two
different renormalization schemes, NDR and HV, with consistent
results.
We give many details on the calculation of the
anomalous dimension matrix at two loops, on the determination
of the Wilson coefficients
at the scale $M_W$ and of their evolution from  $M_W$ to $\mu$.
We also discuss the dependence of the Wilson coefficients/operators
on the regularization scheme.
 \end{abstract}
\date{}
\newpage
\section{Introduction}
\label{sec:intro}
In this paper we present a calculation of the two-loop
anomalous dimension matrix relevant for $\Delta S=1$ decays.
The anomalous dimension matrix includes leading and
sub-leading corrections at orders $\alphas$,
$\alphas^2$, $\alphae$ and $\alphas \alphae$.  The calculation
has been performed in two different regularization schemes:
naive dimensional regularization (NDR) and 't Hooft-Veltman
regularization (HV)\cite{hv}.
We verify that the results obtained in the
two schemes are compatible both in the strong and electro-magnetic
case. We give many details on the calculation itself,
on the definition
of the renormalized operators, on the relation between different
regularization schemes, on the role of counter-terms for operators
which vanish by the equations of motion, on  the gauge invariance
of the final result etc. We also report  a list containing
the double and single
pole contribution of all the diagrams, counter-terms and
``effervescent" operator counter-terms in both the schemes used
in this calculation. The list of all the diagrams
may be useful to check our results and for further applications.
\par A comparison of our results with a parallel calculation
reported in refs.\cite{bjlw1}-\cite{bjl} is also presented.
We agree with the authors of these references in the NDR scheme.
For some diagrams computed in the HV regularization scheme
we however disagree and we explain the origin
of the difference. In the electro-magnetic case we show that, by
using the  values of ref.\cite{bjlw2} for the diagrams computed in HV,
 it is not possible to satisfy
the expected relation between the two-loop anomalous dimension
matrix computed in NDR and HV.
On the contrary our results satisfy the expected relations for
both terms of order $\alphas^2$ and $\alphas \alphae$.
We have also checked the consistency of our calculation
at order $\alphae^2$. \par The authors of ref.\cite{bjl} have only
computed the anomalous dimension in NDR and derived the result
in HV by using the one-loop anomalous dimension and coefficient
matrix. We thus agree with their final result for the
two-loop anomalous dimension matrix in HV too, in spite of
the different results for some diagrams between this work and
ref.\cite{bjlw2}.
\par In this
paper we have preferred to report  only the calculation of
the two-loop anomalous dimension matrix with as many details
as possible and postpone a discussion of the numerical calculation
and uncertainties for the coefficient functions\footnote{
Dependence of the coefficient functions on the
renormalization scale $\mu$, on $\Lambda_{QCD}$ and on the
renormalization prescription for example.}
 to a separate
publication\cite{noin}.
A phenomenological analysis of $\epsilon^{\prime}
/ \epsilon$, using the results reported in this paper,
 has already been presented in ref.\cite{noi} and will be
extended/upgraded in ref.\cite{noin}.
\par
The paper  is organized as follows. In sec.\ref{sec:gefo} we introduce
the general formalism. The calculation of the coefficient functions
at the $W$ scale is summarized in sec.\ref{sec:mc}.
A detailed discussion of the scheme dependence of the anomalous
 dimension matrix  and of the relation between different
renormalization schemes is given in sec.\ref{sec:andim}.
We also present a convenient definition of the renormalized operators
 which makes the evolution matrix scheme independent.
This definition may be useful to predict weak amplitudes by
combining  the Wilson coefficients with the matrix elements
of the corresponding operators computed with a non-perturbative method,
as for example lattice QCD\cite{noin,mpst}.  In the same section we also
recall some basic features of the HV and NDR regularizations.
 Sec.\ref{sec:an3} is the main section of this paper.
 There we describe the calculation of one- and two-loop diagrams,
discuss the role of the so called ``effervescent" operators and
comment on the subtraction of counter-terms corresponding
to operators which vanish
by the equations of motion. Double and single pole contributions
of all the relevant Feynman diagrams, in the NDR and HV regularizations
are given in the Appendix. From the calculation of the
Feynman diagrams, the two-loop anomalous dimension matrix is
derived  and given in the NDR and HV schemes
in sec.\ref{sec:gelettro}.
\section{General Formalism}
\label{sec:gefo}
Effective Hamiltonians for non-leptonic decays
of hadrons composed by light quarks ($K$, $D$ and $B$ mesons
for example)  are defined by Wilson operator expansions of
products of weak currents\cite{gale}-\cite{alma}:
\bea < F \vert {\cal H}_{eff}\vert I > &=& g_W^2/8 \int d^4 x
D_W(x^2, M_W^2) < F \vert T \left( J_{\mu}(x),J^{\dagger}_{\mu} (0) \right)
\vert I > \nn \\ &\rightarrow& \sum_{i}
< F \vert Q_{ i}(\mu)  \vert I > C_{ i}(\mu)  \label{wope} \eea
For kaon decays, in the limit in which we neglect quark masses,
only four-quark operators appear on the r.h.s. of eq.(\ref{wope}).
The $\Delta S=1$ effective hamiltonian can then be written as :
\bea
{\cal H}_{eff}^{\Delta S=1}&=&\lambda_u \frac {G_F} {\sqrt{2}}
\Bigl[ (1 - \tau ) \Bigl( C_1(\mu)\left( Q_1(\mu) - Q_1^c(\mu) \right) +
C_2(\mu)\left( Q_2(\mu) - Q_2^c(\mu) \right)  \Bigr)\nn\\
&+&\tau\sum_{i=1} Q_{ i}(\mu) C_{ i}(\mu) \Bigr]
\label{eh} \eea
where $\lambda_u = V_{ud} V^*_{us}$ and similarly we can define
$\lambda_c$ and $\lambda_t$. $\tau=-\lambda_t/\lambda_u$ and $V_{ij}$ is
one of the elements of the CKM\cite{cab,km} mixing matrix.
The operator basis is given by:
\bea
Q_{ 1}&=&({\bar s}_{\alpha}u_{\beta})_{ (V-A)}
    ({\bar u}_{\beta}d_{\alpha})_{ (V-A)}
   \nn\\
Q_{ 2}&=&({\bar s}_{\alpha}u_{\alpha})_{ (V-A)}
    ({\bar u}_{\beta}d_{\beta})_{ (V-A)}
\nn \\
Q_{ 3,5} &=& ({\bar s}_{\alpha}d_{\alpha})_{ (V-A)}
    \sum_{q=u,d,s,\cdots}({\bar q}_{\beta}q_{\beta})_{ (V\mp A)}
\nn \\
Q_{ 4,6} &=& ({\bar s}_{\alpha}d_{\beta})_{ (V-A)}
    \sum_{q=u,d,s,\cdots}({\bar q}_{\beta}q_{\alpha})_{ (V\mp A)}
\nn \\
Q_{ 7,9} &=& \frac{3}{2}({\bar s}_{\alpha}d_{\alpha})_
    { (V-A)}\sum_{q=u,d,s,\cdots}e_{ q}({\bar q}_{\beta}q_{\beta})_
    { (V\pm A)}
\nn \\
Q_{ 8,10} &=& \frac{3}{2}({\bar s}_{\alpha}d_{\beta})_
    { (V-A)}\sum_{q=u,d,s,\cdots}e_{ q}({\bar q}_{\beta}q_{\alpha})_
    { (V\pm A)} \nn \\
Q^c_{ 1}&=&({\bar s}_{\alpha}c_{\beta})_{ (V-A)}
    ({\bar c}_{\beta}d_{\alpha})_{ (V-A)}
\nn \\
Q^c_{ 2}&=&({\bar s}_{\alpha}c_{\alpha})_{ (V-A)}
    ({\bar c}_{\beta}d_{\beta})_{ (V-A)}
\label{epsilonprime_basis}
\eea
when QCD and QED corrections are taken into
account\cite{gale,alma}, \cite{harla}-\cite{bw}.
In (\ref{epsilonprime_basis})
 the subscript $(V \pm A)$ indicates the chiral structure and
$\alpha$ and $\beta$ are colour indices. The sum is intended over
those flavours which are active at the scale $\mu$.
We have completely
ignored the effects due to the operators $Q_1^b$ and $Q_2^b$
which are the analog of $Q_1^c$ and $Q_2^c$ with the charm quark replaced
by the bottom quark. In ref.\cite{harla}
it was indeed shown that these operators
have a negligible effect on the evolution of the Wilson coefficients
of the operators (\ref{epsilonprime_basis}).  Their inclusion,
once that the anomalous dimension matrix is known, is in any case elementary.
For $m_c < \mu < m_b$ we have used the relation:
\be Q_{10} = Q_9 + Q_4 -Q_3 \ee
to eliminate $Q_{10}$ from the evolution equations.
We think that expression (\ref{eh}) is the most trasparent for $\mu > m_c$.
It shows that we can find all the Wilson coefficients by evolving $C_{1,10}$
($C_{1,9}$)
via a $10 \times 10$  ($9 \times 9$) evolution matrix down to $\mu= m_b$
($m_c < \mu < m_b$). The generalization to $\mu < m_c$ is
straightforward and can be found, for example, in a recent paper on
$\epsilon^{\prime} / \epsilon$\cite{bjln}.

\par The operators $Q_i(\mu)$ are renormalized at the scale $\mu$
in some given scheme.  The corresponding coefficients
$C_i(\mu)$ are scheme dependent. The dependence on the regularization
scheme appears at one loop, when we express the original current-current
product
in terms of the Wilson  operator product expansion (OPE), see eq.(\ref{wope}).
\par To obtain the coefficients $C_i(\mu)$ at next-to-leading order
(NLO) two steps are necessary:
\par
1) The calculation of the coefficients at a given scale, for example
$M_W$ or $m_t$, including corrections of order $\alphas$ and
$\alphae$. \par
2) The calculation of the two-loop anomalous dimension up to
$O(\alphas^2)$ and $O(\alphas \alphae)$\footnote{ Through this paper
we neglect terms of order $\alphae^2$.}. \par
The results of steps 1) and 2) depend on the regularization scheme and on
the normalization conditions imposed on the renormalized operators, as will
be discussed below. We have done our calculations in two popular
regularization schemes, i.e. the t'Hooft-Veltman (HV)\cite{hv}
 and the naive (NDR)
dimensional regularization schemes. In both cases we have obtained the
renormalized operators via the standard modified minimal subtraction
procedure $\overline{MS}$. We will also discuss other renormalization
prescriptions which can make the renormalization group evolution matrix
scheme independent. \par In presence of $\gamma_5$ and in certain
regularization
schemes,  the axial vector current
may develop an anomalous dimension at the two-loop level. In defining the
evolution matrix one has to take into account this effect. Alternatively
one can impose to the
current a certain one-loop renormalization condition
such that its two-loop anomalous dimension is zero. We prefer
this second solution and discuss this point in sec.\ref{sec:mc}. \par
To make easier a comparison with previous calculations on the same
subject, \cite{bjlw1}-\cite{bjl} and \cite{acmp}, we follow as close as
possible the notation introduced in ref.\cite{bjlw1} and write:
\be
{\cal H}_{eff}^{\Delta S=1} \sim  \vec Q^T(\mu) \, \vec C(\mu)
\label{eh1} \ee
where $\vec Q^T(\mu)$ is a row vector whose components are the operators
$Q_{1,10}$ of the basis
(\ref{epsilonprime_basis}) and $\vec C(\mu)$ is a column vector,
whose components are the corresponding Wilson coefficients.
$\vec C(\mu)$ are expressed
in terms of  $\vec C(M_W)$ through the
 renormalization group evolution matrix $\W[\mu,M_W]$:
\be \vec C(\mu) = \W[\mu,M_W] \vec C(M_W) \label{evo} \ee
The coefficients $\vec C(\mu)$ obey the renormalization group equations:
\be
\Bigl( - \frac {\partial} {\partial t} + \beta ( \alphas )
\frac {\partial} {\partial \alphas} - \frac {\hat \gamma ( \alphas ,
\alphae ) }{2} \Bigr) \vec C(t, \alphas(t), \alphae) =0 \label{rge} \ee
where $t=ln ( M_W^2 / \mu^2 )$ and we ignore the running of $\alphae$.
The factor of $2$ in eq.(\ref{rge}) normalizes the anomalous dimension
matrix as in refs.\cite{bjlw1}-\cite{bjl}.
\be \hat \gamma = \hat \gamma_{Q} - 2 \,
\gamma_J \, \hat 1 \label{vattelapesca} \ee
 is the anomalous dimension matrix of the operators minus
twice the anomalous dimension of the weak current in a given renormalization
scheme. In eq.(\ref{vattelapesca}) $\hat 1$ is the identity matrix.
\par To simplify the discussion,
we first consider the case where there is   no crossing of
a quark threshold when going from
 $M_W$ to $\mu$. The relevant formulae for the general
case will be  given at the end of this
section.
At the next-to-leading order,
by expanding $\W[\mu,M_W]$, we can write:
\bea
\W[\mu,M_W]   = \hat M[\mu] \U[\mu, M_W] \hat M^{\prime}[M_W]
 \label{monster} \eea
with:
\be  \hat M[\mu] =
\left(\hat 1 +\frac{\alphae }{4\pi}\Ke\right)
  \left(\hat 1 +\frac{\alphas (\mu)}{4\pi}\J\right)
 \left(\hat 1+\frac{\alphae}{\alphas (\mu)}\PP\right)
\label{mo1} \ee
and
\be \hat M^{\prime}[M_W]=\left(\hat 1-\frac{\alphae}{\alphas (M_W)}\PP\right)
            \left(\hat 1 -\frac{\alphas (M_W)}{4\pi}\J\right)
\left(\hat 1 -\frac{\alphae }{4\pi}\Ke\right)
\label{mo2} \ee
\par We substitute the expression of  $\vec C(\mu)$
given in eq.(\ref{evo}) in the renormalization group equations (\ref{rge}),
using $\hat W[\mu, M_W]$ written
as in eqs.(\ref{monster}-\ref{mo2}).
By expanding  the anomalous dimension matrix, which includes gluon and photon
corrections, up to order $\alphas^2$ and $\alphae \alphas$:
\be \hat \gamma= \frac {\alphas }{ 4 \pi } \hat \gamma_s^{(0)} +
 \frac {\alphae }{4 \pi} \hat \gamma_e^{(0)}
+ (\frac {\alphas }{4 \pi})^2 \hat \gamma_s^{(1)} +
 \frac{ \alphas }{4 \pi} \frac{ \alphae}{4 \pi}  \hat \gamma_e^{(1)}
\nn \ee
we obtain the expression for $\U$:
\bea
\U[\mu,M_W]&= & T_{\alphas} exp \Bigl(- \int_{\alphas(M_W)}^{\alphas(\mu)}
\frac{d\alphas}{\alphas} \frac{\hat \gamma^{(0) T}_s}{2\beta_0} \Bigr)
 \nn \\ &\rightarrow& \left[\frac{\alphas (M_W)}{\alphas (\mu)}\right]^{
            \gammazeros / 2\beta_{ 0}}
\label{u0} \eea
in the basis where $\hat \gamma^{(0)}_s$ is diagonal. $T_{\alphas}$ is the
ordered product, with increasing couplings from right to left.
The matrices $\PP$, $\J$ and $\Ke$ are solutions of the equations\footnote
{ The last term in eq.(15) of ref.\cite{noi}, corresponding
to eq.(\ref{ke})  of this paper, contains a misprint.
The expression  which is reported here has however been used in all
 the numerical
calculations of the Wilson coefficients of ref.\cite{noi}, which are
consequentely correct. We thank A. Buras for finding the misprint.}:
\be
\PP+\left[\PP,\frac{\gammazeros}{2\beta_{ 0}}\right] = \frac{\gammazeroe}
   {2\beta_{ 0}} \label{pp} \ee
\be
\J-\left[\J,\frac{\gammazeros}{2\beta_{ 0}}\right] =
         \frac{\beta_{ 1}}{2\beta^2_{ 0}}\gammazeros-
         \frac{\gammaones}{2\beta_{ 0}} \label{jj}
\ee
\be
\left[\Ke,\gammazeros \right]
=\gammaonee+\gammazeroe \J+\gammaones \PP+\left[\gammazeros,\J\PP\right]
    -2\beta_1 \PP -\frac{\beta_{ 1}}{\beta_{ 0}}\PP\gammazeros
\label{ke} \ee
In eqs. (\ref{u0}-\ref{ke}),  $\beta_{ 0}$ and $\beta_{ 1}$ are the
first two coefficients of the $\beta$-function of $\alphas$.
$\hat U$ and $\PP$ are determined by the leading logarithmic (LO) anomalous
dimension matrices $\hat \gamma^{(0)}_s$ and $\hat \gamma^{(0)}_e$ and
are regularization scheme independent. On the other hand
the two-loop anomalous dimensions $\hat \gamma_s^{(1)}$ and
$\hat \gamma_e^{(1)}$, and consequently $\J$, $\Ke$ and $\hat W[\mu , M_W]$,
 are regularization scheme dependent.
Eqs.(\ref{pp}-\ref{ke}) can be easily solved in the basis where $\hat
\gamma^{(0)}_s$ is diagonal. The solutions develop singularities which however
cancel in the final expression of $\hat W[\mu,M_W]$. It is indeed possible
to find an explicit form of $\hat W[\mu,M_W]$ which is not singular.
This form was used in the numerical calculation of ref.\cite{noi}.
\par The initial conditions for the evolution equations,
$\vec C(M_W)$ are obtained by matching the full theory, which includes
propagating  $W$,
$Z^0$ and six quarks, to the effective theory, where the $W$,
$Z^0$ and top quark have been removed simultaneously. In general,
$\vec C(M_W)$ depend on the definition of the renormalized operators
in a given regularization scheme. A scheme independent way of defining
$\vec C(M_W)$ and the evolution matrix will be discussed in
sec.\ref{sec:andim}.
In the full theory, all the coefficients coming from
current-current and penguin operators have been computed in
refs.\cite{harla},\cite{flynn}-\cite{inami}. In this work we have only
computed, in
the effective theory for the HV and NDR regularization schemes,
the $O(\alphas)$ and $O(\alphae)$ one-loop corrections necessary
to impose the matching conditions on $\vec C(M_W)$.
\par
We now discuss the modifications to the evolution matrix in presence of
a heavy quark threshold.
These modifications are necessary when $\mu < m_b$. The estension to the
case when
$\mu < m_c$ is straightforward. The matrix
$\hat W[\mu_1,\mu_2]$ depends on the number of active flavours $f$ in the
interval $[\mu_1,\mu_2]$.
We denote it by $\hat{W}_f[\mu_1,\mu_2]$.
On the other hand when we cross a threshold, the renormalization conditions
of the operators are in general changed\footnote{This however does not happen
with scheme independent renormalization conditions.}.
At threshold
it is thus necessary to introduce a suitable matrix $\hat T$ \footnote{
The matrix $\hat T$ is different for different thresholds, depending
on the quark electric charge.}.
This matrix allows for the
matching of the evolution between scales larger and smaller than the
threshold.
Thus for example, to evolve the coefficients from $M_W$ to $\mu < m_b$, one
has to use the following expression\cite{sch}:
\be
\hat{W}[\mu,M_W]=\hat{W}_4[\mu,m_b]\hat T  \hat{W}_5[m_b,M_W]
\label{wth}
\ee
where
\be
\hat T= \hat 1+\frac{\alphas(m_b)}{4\pi}\delta\hat{r}^T+\frac{\alphae}
{4\pi}\delta\hat{s}^T
\label{matm}
\ee
The matrices $\delta\hat r$ and $\delta\hat s$ relevant for eq.(\ref{wth})
are given in sec.\ref{sec:oneloopres} together with all other
one-loop results.

\section{ Calculation of the Coefficients $\vec C(M_W)$}
\label{sec:mc}
In order to compute the Wilson coefficients of the OPE at
a scale $\mu \sim M_W$  ($m_t$), we have to consider the full set of
current-current, box and strong, electro-magnetic and $Z^0$ penguin
diagrams up to and including $O(G_F \alphas)$ and
$O(G_F \alphae)$\cite{harla},\cite{flynn}-\cite{inami}.
In the current-current case we compute the diagrams
with external momenta $\vert p_i^2 \vert \sim \mu_0^2 \ll M_W^2$ (i.e.
we neglect terms of order $\mu_0^2 / M_W^2$) with massless
external quark states. The dependence on the external momenta only
appears in logarithms, proportional to the anomalous dimension of
the operators, and in the
operator matrix elements. Strong and electro-magnetic penguin diagrams
are also computed
with external momenta $\vert p_i^2 \vert \ll M_W^2$
($\vert p_i^2 \vert \ll m_t^2$) and
massless external states.  In the case of penguin diagrams,
the logarithmic dependence on the external states appears as a dependence on
the momentum transferred through the gluon or photon propagators,
$q^2 \sim \mu_0^2$.
$Z^0$-penguin and box diagrams can be computed with zero  external momenta
and including only the top quark contribution,
 since they are infrared finite, cf. $B(x_t)$ and $C(x_t)$ in table
\ref{tab:inali}.
\par We now introduce  the notation necessary for
the calculation of the coefficients $\vec C(M_W)$.
In the full theory, the direct calculation of the current-current, box and
penguin diagrams at one loop (including order $\alphas$ and $\alphae$
corrections) has the form:
\bea < J J > &\sim&  < \vec Q^{(0) \, T} > \Bigl[ \vec T^{(0)} +
\frac {\alphas} { 4 \pi} \vec T^{(1)} + \frac {\alphae} { 4 \pi} \vec
D^{(1)}
\Bigr] \nn \\ &=& < \vec Q^T (M_W) > \vec C(M_W)  \label{coe1} \eea
where $< \vec Q^{(0) \, T} > $ are the tree-level matrix elements and
$\vec T^{(1)}$ and $\vec D^{(1)}$ depend
on the regularization scheme and on the external
quark states. Indeed all the $W$-$g$ and $W$-$\gamma$ box diagrams,
being finite, are regularization scheme independent.  The axial vector
vertex diagram however does depend on the regularization scheme.
 The diagrams necessary to obtain $\vec T^{(0),(1)}$ and
$\vec D^{(1)}$ are shown in figs.1,2 and 4-8. By inserting
the renormalized operators of the effective
hamiltonian (\ref{epsilonprime_basis}) in the diagrams
reported in figs.3 and 9, we then compute the
one-loop current-current, strong  and electromagnetic penguin diagrams
between the same external states, using the same regularization
scheme.  In this case we obtain:
\be < \vec Q (M_W) > = \Bigl( 1 +
\frac {\alphas} { 4 \pi} \hat r + \frac {\alphae} { 4 \pi} \hat s
\Bigr) < \vec Q^{(0)} > \label{coe2} \ee  \par
$\vec T^{(1)}$ and $\vec  D^{(1)}$ contain logarithms of the external momenta,
whose coefficients are proportional to the one-loop anomalous dimensions
of the relevant operators (the $W$ and top masses acting as an effective
ultraviolet cutoff). Thus for example $\vec T^{(1)}$ contains terms
$\sim ln (M_W^2 / \mu^2_0)$ ($\sim ln (m_t^2 / \mu^2_0)$).
On the other hand the insertion of the
renormalized operators $\vec Q(\mu)$ between the same external states goes like
$ln (\mu^2 / \mu_0^2)$, where $\mu$ is the renormalization scale.
If we choose the renormalization scale
$\mu = M_W$, the logarithms disappear when we compute the coefficients
$\vec C(M_W)$, which are obtained by comparing eq.(\ref{coe1})
with eq.(\ref{coe2}):
\be \vec C(M_W) = \vec T^{(0)} + \frac {\alphas} { 4 \pi}
\Bigl( \vec T^{(1)} - \hat r^T \vec T^{(0)} \Bigr)  + \frac {\alphae} { 4 \pi}
\Bigl( \vec D^{(1)} - \hat s^T \vec T^{(0)} \Bigr) \label{coe3} \ee
We take $\alphae$ as a fixed coupling constant and $\alphas$ in
eq.(\ref{coe3}) has to be interpreted as $\alphas(M_W)$.
 $\vec T^{(1)}$ and $\hat r^T$ ($\vec D^{(1)}$ and $\hat s^T$)
 depend on the external states and on the regularization scheme. However their
difference depends only on the regularization scheme. We will give the
results for both the HV and NDR regularizations.
\par We now describe
separately the strong and electro-magnetic current-current,
penguin and box diagrams.
\subsection{ Current-Current $O(\alphas)$ Diagrams}
\label{sec:mc1}
The coefficients
$\vec C(M_W)$ depend on the combination $\vec T^{(1)} - \hat r^T \vec
T^{(0)}$ which
is independent of the external states.
Thus we can choose to compute the
diagrams in figs. 2a-2c and 3a-3c with  external states different
from diagram to diagram (but equal  for corresponding
diagrams in the full and effective theory,
 2a and 3a for example). We have
choosen the external momenta as shown in the figures
\footnote{ The same choice of external momenta for the current-current
diagrams has been adopted for the $O(\alphae)$ corrections
described in sec.\ref{sec:mc2}.}. At one loop in
$\alphas$, $< JJ >$ can only mix with the operators $Q_{1,2}$ and
$Q_{1,2}^c$ of the list given above. We only discuss the mixing of $Q_{1,2}$,
since the case $Q_{1,2}^c$ proceeds in the same way.
\par When we compute the vertex corrections to the weak charged current,
fig.2a, and combine it with the renormalization of the external quark lines,
fig.4, the axial current is in general subject to a finite
renormalization which depends on the regularization scheme. A
finite correction to the axial current at one loop
implies that the current have
a non-zero two-loop anomalous dimension. The current anomalous dimension
must be subtracted from the  anomalous dimension of the operators of
the effective Hamiltonian, see eq.(\ref{vattelapesca}). Alternatively
we can apply a finite renormalization to the current in
such a way that  its two--loop anomalous dimension is zero. This
procedure modifies the coefficients $\vec C(M_W)$. It
 is equivalent to impose that the current obeys the Ward identity
which states that, in perturbation theory,
 the axial current must be conserved in the limit in
which the external quark masses vanish.  The two definitions of
the axial current will
give, at the NLO, the same physical result. We prefer to choose
the second alternative,
i.e. $\gamma_J=0$ at two -loops, and avoid the subtraction of the
current anomalous dimension.
\par With our choice of the renormalized weak current, in the
HV scheme, we find:
\bea C_{HV-ccg}^{(1)}(M_W) = \frac{\alphas(M_W)}{4 \pi} \frac{7}{2} \nn \\
C_{HV-ccg}^{(2)}(M_W) = -\frac{\alphas(M_W)}{4 \pi} \frac{7}{6} \eea
where the subscript $ccg$ indicates that this is the contribution
from current-current diagrams, $"cc"$, from the exchange of a gluon,
$"g"$. In the calculation of the coefficient functions $\vec C(M_W)$,
$\alphas(M_W)$ has to be understood as the running coupling constant,
computed in the 5-flavour theory at the scale $M_W$. This is the correct
procedure when $m_t$ is larger than $M_W$, as suggested by the mass lower bound
obtained  for a top quark which decays
in the standard modes.
In the NDR scheme, we find instead:
\bea C_{NDR-ccg}^{(1)}(M_W) = \frac{\alphas(M_W)}{4 \pi} \frac{11}{2} \nn \\
C_{NDR-ccg}^{(2)}(M_W) = -\frac{\alphas(M_W)}{4 \pi} \frac{11}{6} \eea

\subsection{ Current-Current $O(\alphae)$ Diagrams}
\label{sec:mc2}
The electro-magnetic corrections are more complicated due to the presence
of the non-abelian diagrams of figs.5d and 6b and the
$W$-$Z^0$ box diagrams.  The sum of the vertex and self-energy
diagrams, figs.5a and 6, contrary to the $O(\alphas)$ case, is neither finite
nor gauge invariant. This is related to the fact that
the sum of all the diagrams in figs.5 and 6 contributes both to
the renormalization of the Fermi constant $G_F$, which is universal for
quarks and leptons, and to the electromagnetic corrections to the
effective Hamiltonian at order $\alphae$\cite{sirlin1,sirlin2}.
This means that we can reabsorb
a part of the $O(\alphae)$ corrections by a suitable redefinition
of $G_F$.
We proceed following ref.\cite{sirlin2}. In the vertex and self-energy
diagrams in figs.5a and 6a, we write the photon contribution using the
identity:
\be k^{-2}= k^{-2} M_W^2 (M_W^2 - k^2)^{-1} + ( k^2 - M_W^2 )^{-1}
\label{furbo} \ee
The contributions coming from the second term in eq.(\ref{furbo}) are
divergent. The divergent part of these diagrams,
combined with  the divergent terms from diagrams in figs.5d and 6b, gives
 a universal contribution
(equal for quarks and leptons)
to the renormalization of the weak coupling constant $g_W$ and $W$ mass.
Thus this term
 can be reabsorbed in the definition of the physical $G_F$ as measured
in $\mu$-decays\cite{sirlin2}.
The first term in eq.(\ref{furbo})
 is cut-off by the additional convergent factor
$M_W^2 (M_W^2 - k^2)^{-1}$. For this reason it
can only give a finite
contribution  $\sim ln (M_W^2/\mu_0^2)$,  similar to
the contribution of the $\gamma$-$W$ box diagrams of figs.5b-c.
 The sum of the terms
$\sim ln (M_W^2/\mu_0^2)$ gives indeed the current-current
contribution to $\gamma^{(0)}_e$.
The remaining genuine  $O(G_F \alphae)$  corrections
(i.e. not
containing  $ln (M_W^2/\mu_0^2)$ ) are all proportional to
$Q_2$  both in the effective and in the full
theory\cite{sirlin1}-\cite{sirlin2}.
They give thus a correction of order $\alphae / 4 \pi \sim 10^{-3}$
to a coefficient of $O(1)$ and can be safely ignored.
In ref.\cite{bjln} they have instead computed the
finite $O(\alphae)$
corrections by analogy with the $O(\alphas)$ case, i.e. by considering
the differences between the diagrams of fig.3a-c and 5a-c with
a photon exchanged. In this case one would have found:
\bea C_{HV-cc\gamma}^{(2)}(M_W) = -\frac{\alphae}{4 \pi} \frac{13}{6} \nn \\
C_{NDR-cc\gamma}^{(2)}(M_W) = -\frac{\alphae}{4 \pi} \frac{35}{18}
\label{machi} \eea
We observe that the above coefficients do not exaust the $O(\alphae)$
corrections.
For example they do not contain the terms coming from the $Z^0$-$W$ box
diagrams which exist only in the full theory.

\subsection{ QCD Penguin Diagrams}
\label{sec:mc3}
Penguin diagrams have been extensively studied in the literature
\cite{inami}-\cite{harla}.
In the full theory,
for $\mu_0 \gg m_c$, using the unitarity of the CKM
mixing matrix, we have only to consider the contribution coming from
the difference between the top-penguin and the up-penguin diagrams.
The analogous contribution from the charm-up penguins is cancelled
by the GIM mechanism. We can consider the up-quark as massless in the
calculation.

\begin{table}
\begin{center}
\begin{tabular}{|cc|}  \hline\hline
\tsq{\it Inami-Lim Functions} & {\it Expression} \\ \hline
\tsq $B(\xt)$ & $\frac{1}{4}\left[\xt/(1-\xt)+\xt\ln\xt/(\xt-1)^2
\right]$ \\[.5ex] \hline
\tsq$C(\xt)$&$\frac{1}{8}\xt\left[(\xt-6)/(\xt-1)+(3\xt+2)/(\xt-1)^2
 \ln\xt \right] $ \\[.5ex] \hline
\tsq$D(\xt)$&$-\frac{4}{9}\ln \xt+(-19\xt^3+25\xt^2)/[36(\xt-1)^3]+$\\
& $[\xt^2(5\xt^2-2\xt-6)]/[18(\xt-1)^4]\ln\xt $\\[.5ex] \hline
\tsq$E(\xt)$ &$-\frac{2}{3}\ln\xt+[\xt^2(15-16\xt+4\xt^2)]/[6(1-\xt)^4]
\ln\xt+$\\
& $[\xt(18-11\xt-\xt^2)]/[12(1-\xt)^3]$\\[0.5ex]
\hline\hline
\end{tabular}
\caption[]{Basic functions governing the $\mt$-dependence of various weak
amplitudes}
\label{tab:inali}
\end{center}
\end{table}
{}From the diagrams in fig.8a and 9, in the HV scheme, we find:
\bea
C^{(3)}_{HV-pg}(\Mw)\!&=&\!C^{(5)}_{HV-pg}
(\Mw)=-\frac{\alphas(M_W)}{24 \pi} E(\xt)\\
C^{(4)}_{HV-pg}(\Mw)\!&=&\!C^{(6)}_{HV-pg}
(\Mw)=\frac{\alphas(M_W)}{8 \pi}E(\xt) \nn
\eea
where $x_t= m_t^2/M_W^2$. The modified
Inami-Lim function $E(x_t)$ is given in
table \ref{tab:inali}.
In the NDR scheme one finds:
\bea
C^{(3)}_{NDR-pg}(\Mw)\!&=& C^{(5)}_{NDR-pg}
(\Mw)=-\frac{\alphas(M_W)}{24 \pi} \left( E(\xt)\ - \frac{2}{3}\right) \\
C^{(4)}_{NDR-pg}(\Mw)\!&=& C^{(6)}_{NDR-pg}
(\Mw)=\frac{\alphas(M_W)}{8 \pi} \left( E(\xt) -\frac{2}{3} \right) \nn
\eea
These results were already presented in ref.\cite{bjlw1}.
\subsection{ QED Penguin Diagrams}
\label{sec:mc4}
In the case of electro-magnetic penguins we have also to consider,
besides the diagram of fig.8a, the non-abelian diagram given in fig.8b.
In HV, we find the following global contribution for QED-penguin diagrams:
\bea
C^{(7)}_{HV-p\gamma}(\Mw)\!&=&\frac {\alphae}{6\pi}
 D(\xt) \\
C^{(9)}_{HV-p\gamma}(\Mw)\!&=& \frac {\alphae}{6\pi}
D(\xt) \nn
\eea
where the Inami-Lim function $D(\xt)$ is given in table \ref{tab:inali}.
In NDR one finds:
\bea
C^{(7)}_{NDR-p\gamma}(\Mw)\!&=&
\frac{\alphae}{6\pi}\left(
 D(\xt)-\frac{4}{9} \right) \\
C^{(9)}_{NDR-p\gamma}(\Mw)\!&=&
\frac{\alphae}{6\pi} \left(
 D(\xt) -\frac{4}{9} \right) \nn
\eea

\subsection{ $Z^0$ Penguin Diagrams}
\label{sec:mc5}

$Z^0$ penguin diagrams give $O(\alphae)$ contributions\cite{flynn}
 as it was the case
for  photon-penguins. The corresponding modified
Inami-Lim function \cite{inami} vanishes as
$x_t \rightarrow 0$. We can then reasonably
neglect the up and charm quark contributions, even though the
GIM mechanism is not active in this case.
One gets:
\bea C^{(3)}_{HV-pZ^0}(\Mw)\!&=&\frac{\alphae}{6\pi}\frac{1}
{\sin^{ 2}\vartheta_{ W}}
 C(\xt) \\
C^{(7)}_{HV-pZ^0}(\Mw)\!&=&\frac{\alphae}{6\pi}
 4C(\xt) \\
C^{(9)}_{HV-pZ^0}(\Mw)\!&=&\frac{\alphae}{6\pi}
 \left\{4C(\xt)-\frac{1}{\sin^{ 2}\vartheta_{ W}}
 4C(\xt)\right\} \nn
\eea
and the same in NDR.
$C(\xt)$ is given in table \ref{tab:inali}.

\subsection{ Box Diagrams}
\label{sec:mc6}

In the approximation of massless light quarks, the
box diagrams of fig.7 give:
\bea
C^{ (3)}_{(\Box)}(\Mw) &=& \frac{\alphae}{6\pi}
 \frac{1}{\sin^{ 2}\vartheta_{ W}} 2 B(\xt) \\
C^{ (9)}_{(\Box)}(\Mw) &=& \frac{\alphae}{6\pi}
 \frac{1}{\sin^{ 2}\vartheta_{ W}}10 B(\xt) \nn
\eea
both in HV and in NDR.

A summary of the results discussed in this
section for the HV and NDR regularizations is reported
below. We have also included
the tree-level contribution from the diagram of fig.1 and neglected terms
of $O(\alphae / 4 \pi)$ in $C_2$.
\bea  i)\,\,\,\,\, {\it HV} \nn \\
C_{ 1}(\Mw)\!&=&\!\frac{\alphas^{ (5)}(\Mw)}{4\pi}
\frac{7}{2} \nn \\
C_{ 2}(\Mw)\!&=&\!1 - \frac{\alphas^{ (5)}(\Mw)}{4\pi}
\frac{7}{6}\nn \\
C_{ 3}(\Mw)\!&=&\!-\frac{\alphas^{ (5)}(\Mw)}{24\pi}
E(\xt)+\frac{\alphae}{6\pi}\frac{1}{\sin^{ 2}
\vartheta_{ W}}\left[2B(\xt)+C(\xt)\right] \nn \\
C_{ 4}(\Mw)\!&=&\!\frac{\alphas^{ (5)}(\Mw)}{8\pi}
E(\xt) \nn \\
C_{ 5}(\Mw)\!&=&\!-\frac{\alphas^{ (5)}(\Mw)}{24\pi}
E(\xt) \\
C_{ 6}(\Mw)\!&=&\!\frac{\alphas^{ (5)}(\Mw)}{8\pi}
E(\xt) \nn \\
C_{ 7}(\Mw)\!&=&\!\frac{\alphae}{6\pi}\left[
4C(\xt)+D(\xt)\right] \nn \\
C_{ 8}(\Mw)\!&=&\!0 \nn \\
C_{ 9}(\Mw)\!&=&\!\frac{\alphae}{6\pi}\left[
4C(\xt)+D(\xt)+\frac{1}{\sin^{ 2}\vartheta_{ W}}
\Bigl(10B(\xt)-4C(\xt)\Bigr)\right] \nn \\
C_{ 10}(\Mw)\!&=&\!0 \nn
\label{ic_mw_hv}
\eea
and:
\bea  ii)\,\,\,\,\, {\it NDR} \nn \\
C_{ 1}(\Mw)\!&=&\!\frac{\alphas^{ (5)}(\Mw)}{4\pi}
\frac{11}{2} \nn \\
C_{ 2}(\Mw)\!&=&\!1 - \frac{\alphas^{ (5)}(\Mw)}{4\pi}
\frac{11}{6}\nn \\
C_{ 3}(\Mw)\!&=&\!-\frac{\alphas^{ (5)}(\Mw)}{24\pi}
\left( E(\xt) -\frac{2}{3} \right) +\frac{\alphae}{6\pi}\frac{1}{\sin^{ 2}
\vartheta_{ W}}\left[2B(\xt)+C(\xt)\right] \nn \\
C_{ 4}(\Mw)\!&=&\!\frac{\alphas^{ (5)}(\Mw)}{8\pi}
\left( E(\xt) -\frac{2}{3} \right) \nn \\
C_{ 5}(\Mw)\!&=&\!-\frac{\alphas^{ (5)}(\Mw)}{24\pi}
\left( E(\xt) -\frac{2}{3} \right) \\
C_{ 6}(\Mw)\!&=&\!\frac{\alphas^{ (5)}(\Mw)}{8\pi}
\left( E(\xt) -\frac{2}{3} \right)  \nn \\
C_{ 7}(\Mw)\!&=&\!\frac{\alphae}{6\pi}\left[
4C(\xt)+\left( D(\xt) -\frac{4}{9} \right) \right] \nn \\
C_{ 8}(\Mw)\!&=&\!0 \nn \\
C_{ 9}(\Mw)\!&=&\!\frac{\alphae}{6\pi}\left[
4C(\xt)+\left( D(\xt) -\frac{4}{9} \right) +\frac{1}{\sin^{ 2}\vartheta_{ W}}
\Bigl(10B(\xt)-4C(\xt)\Bigr)\right] \nn \\
C_{ 10}(\Mw)\!&=&\!0 \nn
\label{ic_mw_ndr}
\eea

\section{Anomalous Dimensions at One and Two Loops}
\label{sec:andim}
In this section we introduce the notation necessary for the calculation
of the anomalous dimension matrix $\hat \gamma(\alphas , \alphae)$
in dimensional regularization, eq.(\ref{vattelapesca}),
and recall the
rules for the HV and NDR schemes.
\subsection{ General Definitions and Scheme Dependence}
\label{sec:an1}

For simplicity, we start by considering only one- and two--loop corrections
due to strong interactions. The modifications, necessary to include
the electro-magnetic corrections will be given in
sec.\ref{sec:gelettro}
 \par
The anomalous dimension matrix for the operators appearing in the effective
Hamiltonian is defined from the operator renormalization matrix:
\be
\hat\gamma_Q(\alphas)=2 \, \Z^{-1}\mu^2\frac{d}{d\mu^2}\Z
\label{gammag4}
\ee
where $\Z$ is defined by the relation:
\be \vec Q = \Z^{-1} \vec Q^B \label{zd} \ee
which gives the renormalized operators in terms of the bare ones.
\par In a dimensional regularization, as in the HV and NDR schemes,
from eq.(\ref{gammag4}), we obtain:
\be
\hat\gamma_Q = 2 \, \Z^{ -1}\left(-
\epsilon \alphas +\beta(\alphas)\right)\frac{\partial}{\partial \alphas}\Z
\label{RGgamma}
\ee
By writing $\hat\gamma_Q$ and
 $\Z$ as series in the strong coupling constant\footnote{ Since
we normalize the weak currennt in such a way that $\gamma_J=0$,
see below, we have $\hat \gamma= \hat \gamma_Q$.}:
\be \hat\gamma_Q =\frac{\alphas}{4\pi} \hat\gamma^{(0)}_s
 +\frac{\alphas^{ 2}}{(4\pi)^{ 2}} \hat\gamma^{(1)}_s
+ \cdots \label{gexp}\ee
\be
\Z=1+\frac{\alphas}{4\pi}\Z^{ (1)}+
\frac{\alphas^{2}}{(4\pi)^2}\Z^{ (2)}+\cdots
\label{Zexpansion}
\ee
 we  derive the following relations:
\be
\hat\gamma^{ (0)}_s=-2\epsilon\Z^{ (1)}
\label{gamma0}
\ee
and
\be
\hat\gamma^{ (1)}_s  =-4\epsilon\Z^{ (2)}-2 \beta_0\Z^{ (1)}+
2\epsilon\Z^{ (1)}\Z^{ (1)}
\label{gamma1}
\ee
where $\epsilon= (4-D)/2$.
$\beta_0$ is the one-loop coefficient of the $\beta$-function $\beta(\alphas)$
which governs the evolution of the effective coupling constant:
\be \mu^2 \frac { d \alphas} {d \mu^2} = \beta (\alphas ) \label{rcc}\ee
and:
\be \beta(\alphas) = - \beta_0 \frac {\alphas^2} {4\pi}
- \beta_1 \frac {\alphas^3} {(4\pi)^2} + O(\alphas^4) \ee
$\beta_0$ and $\beta_1$ are given by:
\bea \beta_0 &=&\frac {(11N-2 f)} {3} \nn \\
\beta_1 &=& \frac {34}{3} N^2 - \frac{10}{3} N f -\frac
{(N^2-1)}{N} f \eea
where $f$ is the number of flavours.
The running coupling constant, solution of eq.(\ref{rcc}),
is :
\bea \frac {\alphas(\mu^2)}{4 \pi} =
\frac {1} {\beta_0 ln(\mu^2/\Lambda_{QCD}^2)} \Bigl(
1 - \frac{\beta_1 ln[ln(\mu^2/\Lambda_{QCD}^2)]}{\beta_0^2
ln(\mu^2/\Lambda_{QCD}^2)}\Bigr) + \cdots \label{srcc} \eea
The above equation defines $\Lambda_{QCD}$ at the NLO. \par
 We can expand $\Z^{(i)}$ in eqs.(\ref{gamma0}) and (\ref{gamma1})
in inverse powers of $\epsilon$:
\be
\Z^{ (i)}=\sum_{j=0}^{i}\left(\frac{1}{\epsilon}\right)^{ j}
\Z^{ (i)}_{ j}
\label{Zpoleexp}
\ee
The anomalous dimension is finite as $\epsilon \rightarrow
0$. This implies a relation between the one- and two- loop
coefficients of $\Z$:
\be
4\Z^{(2)}_{ 2}+2 \beta_0 \Z^{ (1)}_{ 1}
-2\Z^{ (1)}_{ 1}\Z^{ (1)}_{ 1}=0
\label{twoloop_ad_condition}
\ee

{}From the above equations we finally  obtain:
\be \hat \gamma^{(0)}_s= -2 \Z^{(1)}_1 \label{1loopADM} \ee
and
\be
\hat\gamma^{ (1)}_s=-4\Z^{ (2)}_{ 1}-2 \beta_0 \Z_{
0}^{ (1)}+2(\Z^{ (1)}_{ 1}\Z^{ (1)}_{ 0}+
\Z^{ (1)}_{ 0}\Z^{ (1)}_{ 1})
\label{2loopADM}
\ee
We thus conclude that it is sufficient to compute the pole and
finite  part of $\Z^{(1)}$ and the single pole of $\Z^{(2)}$
in order to obtain the two-loop anomalous dimension.
Eq.(\ref{2loopADM}) tells us how to derive $\hat \gamma^{(1)}$.
In dimensional regularizations, such as HV, NDR or DRED (dimensional
reduction) however, the calculation is complicated by the
presence of the so called "effervescent" operators (EO),
 which appear in
the intermediate steps of the calculation \cite{acmp,bw}.
The EO are independent operators which are present in D-dimensions
but disappear in the physical basis of operators in 4-dimensions.
Because of the presence of the EO, the products of the matrices
$\Z^{(i)}_j$ in eq.(\ref{2loopADM}) have to be done by summming indices
over the full set of operators, including the EO. Only
at the end of the calculation we can  restrict the set of operators
to the operators of the physical 4-dimensional basis
(\ref{epsilonprime_basis}).
 \par Different renormalization prescriptions
will define different renormalized operators. This happens for example
if we adopt the $\overline{MS}$ subtraction procedure in the HV and
NDR schemes. Let us denote as $\Z_{HV}$ ($\Z_{NDR}$) the
renormalization matrices in the two cases and
$\hat r_{HV}$ ($\hat r_{NDR}$), cf. eq.(\ref{coe2}),
the one-loop matrices of the effective  theory in the two regularizations. Then
we must have:
\be \Z_{HV} =\Z_{NDR} \Bigl( \hat 1 + \frac { \alphas} {4\pi}
\Delta \hat r \Bigr) \label{zdiff} \ee
with $\Delta \hat r = \hat r_{NDR} - \hat r_{HV}$.
Using eq.(\ref{gammag4}), expanded in powers of $\alphas$ as in
eqs.(\ref{gexp})-(\ref{Zexpansion}), we then find\footnote{
The corresponding formula for $\hat\gamma^{(1)}_e$ is $
\Delta \hat \gamma^{(1)}_e = \Bigl[ \Delta \hat r ,\hat \gamma^{(0)}_e
\Bigr] +
\Bigl[ \Delta \hat s ,\hat \gamma^{(0)}_s
\Bigr]$.}:
\be \Delta \hat \gamma^{(1)}_s = \Bigl[ \Delta \hat r ,\hat \gamma^{(0)}_s
\Bigr] + 2 \beta_0 \Delta \hat r \label{gdiff} \ee
Eq.(\ref{gdiff}) implies that the combination:
\be \hat G = \hat \gamma^{(1)}_s - \Bigl[ \hat r , \hat \gamma^
{(0)}_s \Bigr] - 2 \beta_0 \hat r \label{hr} \ee
is regularization independent.  It can be shown
that a consequence of eq.(\ref{hr}) is the scheme independence of
the combination $\hat R =\hat r^T + \J$ ($\hat S=\hat s^T + \Ke$
in the case of electromagnetic corrections) on the regularization
scheme, see for example ref.\cite{bjlw1}.
$\hat R$ and $\hat S$ are precisely the combinations which
appear in the final expressions of the coefficient functions\footnote{
This is strictly true if we include the terms of $O(\alphae)$ of
eq.(\ref{machi}).}.
 \par
Using eqs.(\ref{monster})-(\ref{mo2}) and (\ref{coe3}), we obtain:
\be \vec C(\mu) = \hat M[\mu] \hat U[\mu,M_W] \hat N^{\prime}[M_W]
 {\vec C}'(M_W)  \label{cc} \ee
where $\hat M[\mu]$ has been defined in eq.(\ref{mo1}) and:
\be \hat N^{\prime}[M_W]=
\left(\hat 1-\frac{\alphae}{\alphas (M_W)}\PP\right)
 \left(\hat 1 -\frac{\alphas (M_W)}{4\pi}
[ \hat r^T + \J ] \right)
\left(\hat 1 -\frac{\alphae }{4\pi}[ \hat s^T + \Ke ] \right)
\label{rir} \ee
\be \vec C'(M_W) = \vec T^{(0)} + \frac {\alphas} { 4 \pi}
 \vec T^{(1)}  + \frac {\alphae} { 4 \pi}
 \vec D^{(1)}
\label{rir1} \ee
In the above equation we have neglected higher order terms
in $\alphas$ or $\alphae$. From eq.(\ref{rir}),
 we conclude that the matrix $\hat N^{\prime}[M_W]$
is independent
of the regularization. We can obtain scheme independent coefficients
$\vec C(\mu)$ by a suitable redefinition of the renormalized
operators $\vec Q^T (\mu)$:
\be {\vec V}^T(\mu)= \vec Q^T(\mu) \left( 1 - \frac {\alphas
(\mu)} {4 \pi} \hat r^T - \frac {\alphae} {4 \pi} \hat s^T \right)
\ee
With the above redefinition, one has:
 \be  \hat M[\mu] \rightarrow \hat N[\mu] =
\left(\hat 1 +\frac{\alphae }{4\pi} ( \Ke + \hat s^T)\right)
  \left(\hat 1 +\frac{\alphas (\mu)}{4\pi}(\J +\hat r^T) \right)
 \left(\hat 1+\frac{\alphae}{\alphas (\mu)}\PP\right)
\label{ml} \ee
so that $\vec C(\mu) = {\hat N}[\mu] \hat U[\mu, M_W]
\hat N^{\prime}[M_W] {\vec C}'(M_W)$ is regularization
scheme independent. \par
It is not difficult to understand how it is possible to find
 renormalization conditions which do not depend on the scheme.
Let us fix the renormalization conditions by
imposing that the matrix elements of the
renormalized operators have a given value
for a certain set of external quark  (gluon) states:
\be < \beta \vert \vec V \vert \alpha > = 1 \label{ab}\ee
This procedure defines the same renormalized operators,
i.e. the same coefficients $\vec C(\mu)$, in all the
regularization schemes\footnote{ If the renormalized strong coupling
 constant differs in two different renormalization schemes,
as it is the case for HV and DRED for example, the coefficients
will have the same expression only when given in terms of the
same renormalized $\alphas$}. Notice however that the coefficient
functions depend now on the external states chosen to fix
the renormalization conditions, $\vert \alpha >$ and
$\vert \beta >$ in eq.(\ref{ab}). The external states have thus to be
specified if one wants to use renormalization scheme independent
coefficients.  The values of the coefficients can change
in a substantial
way by going from the $\overline{MS}$ HV or NDR prescription
to the scheme independent one. For example, using $\overline{MS}$ HV
we found that $C_6$ is reduced by the inclusion of the NLO
corrections\cite{noi}
while it is enhanced in the scheme independent case\cite{bjlw1,noin}.
We remark that a consistent treatment of the Wilson coefficients
and  renormalized operators is necessary in order to get, up
to higher order corrections, the physical result. Such a treatment
is possible, at least in principle, in lattice QCD but not
in other approaches as for example the $1/N$ expansion.
 \par
Expression (\ref{2loopADM}), which allows us to compute the two-loop
anomalous dimension in terms of the one- and two- loop renormalization
matrices, is indeed valid diagram by diagram. This means that
 we can define the contribution of any given two-loop diagram to
the anomalous dimension by combining:
\par i) the single pole obtained
from the diagram where we insert a given bare operator;
\par ii) the single pole of the diagram obtained by substituting
to any divergent sub-diagram the
appropriate counter-terms, including those proportional to
``effervescent" operators;
\par iii) the single pole coming from the
substitution, in any divergent sub-diagram,
of an appropriate combination of effervescent operators
to take into account  the term
$2(\Z^{ (1)}_{ 1}\Z^{ (1)}_{ 0}+\Z^{ (1)}_{ 0}\Z^{ (1)}_{ 1})$
in eq.(\ref{2loopADM}).
\par  We will call the contribution to
the anomalous dimension of a given diagram plus the counter-terms and
the insertion of the effervescent operators, i)-iii),  the ``complete"
contribution. The advantage of combining different terms
  (bare diagram, counter-terms and insertion of effervescent operators)
diagram by diagram is that this allows several checks on
the contribution of any given two-loop diagram to the anomalous
 dimension. Thus for example, in HV, two diagrams which go one into
another via a Fierz rearrangement, diagrams $V_{10}$ and $V_{12}$
in fig.10 or $P_2$ and $F_2$ in fig.11
for instance, give the same ``complete"
contribution to $\hat \gamma^{(1)}$ for left-left operators\cite{bw}.
\par Similarly, relation (\ref{gdiff}) can be shown to be true for the
``complete"
contribution of any single two-loop diagram. It is  then possible to use it as
  a further check of the calculation in two different regularization schemes.
Eq.(\ref{gdiff}), used diagram by diagram, is a relation
which connects single poles and finite terms at one loop with
single poles at two loops. When satisfied, it authomatically ensures
that the relation holds both for $\alphas^2$ and $\alphae \alphas$
corrections.
A more extensive discussion of the checks done on the calculation
of the single diagrams will be given in sections \ref{sec:2lcc}
and \ref{sec:2lp}. Before this
we  recall some basic facts about the
HV and NDR regularization schemes, sec.\ref{sec:an2},
 and  introduce the ''effervescent" operators by considering the one-loop
diagrams in these two regularizations, secs.\ref{sec:an4}-\ref{sec:an6}.
Then we explain the construction of ``complete"
diagrams by two specific examples, one taken from the set
of current-current diagrams, fig.10, and the other from the set
of penguin diagrams, fig.11, sec.\ref{sec:2lcc} and \ref{sec:2lp}.
\subsection{HV and NDR Regularization Schemes}
\label{sec:an2}
In this subsection we report for completeness the computational
rules of the HV and NDR regularization schemes, which have been
used in the present work. In both the schemes
Feynman diagrams are regularized by performing the integrals
 over the loop momenta in $D=4-2 \epsilon$ dimensions. The
two schemes differ in the definition of $\gamma_5$,
which, in the case of the NDR regularization, can lead to
inconsistencies and has to be treated with particular care.
\begin{enumerate}
\item {\it Naive Dimensional Regularization} (NDR)

In NDR $\gamma$-matrices are $D$-dimensional and only the
$D$-dimensional
metric tensor is introduced, following the convention:
\be
\phantom{mm} g_{\mu\nu}=g_{\nu\mu}\phantom{m} , \phantom{mm} g_{\mu}^{\rho}
g_{\rho}^{\nu}=g_{\mu}^{\nu}\phantom{m} , \phantom{mm}
g_{\mu}^{\mu}=D\phantom{m} , \phantom{mm} Tr(\iii)=4
\label{gd} \ee
The $D$-dimensional Dirac matrices obey the usual algebra:
\be
\{\gamma_{\mu},\gamma_{\nu}\}=2g_{\mu\nu}
\ee
and $\Gfive$ anticommutes with all the $\gamma$'s:
\be
\{\Gfive,\gamma_{\mu}\}=0
\ee
The above definition of $\gamma_5$ may give problems in the evaluation
of closed odd parity fermion loops, as for example $Tr(\Gfive
\gamma_{\mu} \gamma_{\nu} \gamma_{\rho} \gamma_{\sigma})$, which
are not unambiguously defined. For this reason it is not
garanteed that it is possible to obtain the correct results by
using the  NDR scheme. In ref.\cite{bjlw2} and in the present
calculation it is shown that, up to two loops,
 by fixing the ambiguity in the
closed odd parity fermion loops, it is possible to obtain results
which are in agreement with those obtained in the HV regularization
scheme. In the present work we have fixed the ambiguity following the
prescription of ref.\cite{schoon}. We have verified that the prescription
of ref.\cite{schoon} gives the correct result for the triangle anomaly,
as it is also the case for the HV scheme.

\item {\it t'Hooft-Veltman Regularization} (HV)

The HV regularization scheme has been proved to be unaffected
by ambiguities or
inconsistencies in the algebra of the $\gamma$ matrices
\cite{hv,akyea,briet}.
The $D$-dimensional $\gamma$-matrices
$\gamma^{\mu}$ are decomposed in two parts, a 4-dimensional part
$\tilde\gamma^{\mu}$ and a $(D\!-\!4)$-dimensional part $\hat\gamma^
{\mu}$, i.e.:
\be
\gamma^{\mu}=\tilde\gamma^{\mu}+\hat\gamma^{\mu}
\ee
with different anticommutation relations
 with respect to $\Gfive$. The 4-dimensional $\gamma$'s anticommute with
$\Gfive$, while the
$\hat \gamma$'s commute with $\Gfive$:
\be
\{\tilde\gamma_{\mu},\Gfive\}=0\,\,\,\,\,,\,\,\,\,\,
[\hat\gamma_{\mu},\Gfive]=0
\ee
Besides the D-dimensional tensor of eq.(\ref{gd}),
two other metric tensors  are
introduced, namely
 the 4-dimensional tensor  ${\tilde g}_{\mu\nu}$ and
the $(-2\epsilon)
$-dimensional one ${\hat g}_{\mu\nu}$:
\be
\{\tilde\gamma_{\mu},\tilde\gamma_{\nu}\}=2{\tilde g}_{\mu\nu}\,
\,\,\,\,\,,\,\,\,\,\,
\{\hat\gamma_{\mu},\hat\gamma_{\nu}\}=2{\hat g}_{\mu\nu}
\,\,\,\,\,,\,\,\,\,\,
\{\hat\gamma_{\mu},\tilde\gamma_{\nu}\}=0
\ee
with the following ``mixed'' contraction properties:
\be
{\tilde g}_{\mu}^{\mu}=4\,\,,\,\,\,\,
{\hat g}_{\mu}^{\mu}=-2\epsilon\,\,,\,\,\,\,
{\tilde g}_{\mu}^{\rho}{\hat g}_{\rho}^{\nu}=0
\ee
and
\be
{\tilde g}_{\mu}^{\rho}g_{\rho}^{\nu}={\tilde g}_{\mu}^{\nu}
\,\,\,\,\,,\,\,\,\,\,
{\hat g}_{\mu}^{\rho}g_{\rho}^{\nu}={\hat g}_{\mu}^{\nu}
\ee Using HV for
computing Feynman diagrams, it is convenient
to take the external momenta in four dimensions (the loop
momenta being integrated in D-dimension).
\par The bare operators inserted in the one- and two-loop diagrams
are defined by using only 4-dimensional gamma matrices, i.e.:
\be
\tilde\Gmuup(1\!-\!\Gfive)\otimes\tilde\Gmudw(1\!-\!\Gfive)\,\,\,\,\,
\mbox{or}\,\,\,\,\,\tilde\Gmuup(1\!-\!\Gfive)\otimes\tilde\Gmudw(1\!+\!\Gfive)
\label{ophv} \ee
In the algebraic programs, written for the calculation
of the diagrams, it is convenient to write
the weak four-fermion operators in (\ref{ophv}) using
$(1 \pm \Gfive )$ as projectors:
\be
\frac{1}{4}(1\!+\!\Gfive)\Gmuup(1\!-\!\Gfive)\otimes(1\!+\!\Gfive)\Gmudw
(1\!-\!\Gfive)
\ee
and analogously for the $\LR$ case.

\end{enumerate}
\par In 4-dimensions, in the massless theory,
all operators appearing in the weak effective Hamiltonian
are suitable flavour combinations of four-fermion
left-left and left-right operators. This happens because long strings
of $\gamma$ matrices,
appearing at one- or two- loops, can be reduced to left-left
or left-right four-fermion operators, using the completeness
of the Dirac algebra.
 This completeness is lost however using a dimensional regularization.
In this case we can define an infinite
number of independent four-fermion operators which, by quantum numbers,
can mix with the original basis, at different orders in perturbation
theory. The extra operators are those called previously
``effervescent" operators. They do not exist in four dimensions and
are artefact of the regularization. It would be not correct however
to compute the renormalization of the operators without taking
into account the mixing of EO with the 4-dimensional operators.
The reduction to the four dimensional basis can only be done at the
end of the calculation of the two-loop anomalous dimension, as
explained in the next section.
\pagebreak
\section{Diagrams and Counter-terms}
\label{sec:an3}

In this section we report the calculation of the
one-loop diagrams in HV and NDR. This allows us to introduce
the EO induced by the regularization and give the
elements necessary to compute the coefficient functions
$\vec C(M_W)$ as explained in sec.\ref{sec:mc}.
We also give some specific examples of the calculation of
two-loop ``complete" diagrams, both
in the current-current and in the penguin
case, in order to explain the method used in this work and the relations
between the HV and NDR  regularizations. A complete list of
the contribution of all the diagrams is reported in Appendix.
\subsection{Current-Current Diagrams at One Loop}
\label{sec:an4}
Let us consider the diagrams in figs.3a-c,  where
the wavy lines can be due to a gluon or a photon exchange.
The Dirac structure
inserted in the vertex can be of the form
$\LL$ or $\LR$ and we consider both cases separately.
After the loop integration, but before the simplification
of  the Dirac algebra, the pole terms of the diagrams have the following
structure (for the two-loop anomalous dimension we
only need to consider the pole contribution of 4-dimensional
and effervescent operators):
\bea
G^{ (1)}_{ LL,LR}&=&C^{ (1)}\frac{1}{4\epsilon}\left(
\Gnuup\Group\GLup\Grodw\Gnudw\otimes\GLRdw+\GLup\otimes\Gnuup\Group\GLRdw
\Grodw\Gnudw\right)
\nn\\
G^{ (2)}_{ LL,LR}&=&-C^{ (2)}\frac{1}{4\epsilon}\left(
\Gnuup\Group\GLup\otimes\Gnudw\Grodw\GLRdw+\GLup\Gnuup\Group\otimes\GLRdw
\Gnudw\Grodw\right)\nn
\\
G^{ (3)}_{ LL,LR}&=&C^{ (3)}\frac{1}{4\epsilon}\left(
\Gnuup\Group\GLup\otimes\GLRdw\Grodw\Gnudw+\GLup\Group\Gnuup\otimes\Gnudw
\Grodw\GLRdw\right)
\label{vertexCT_1}
\eea
where $C^{ (1)}$, $C^{ (2)}$ and $C^{ (3)}$ are coefficients
which summarize the colour/charge dependence of each diagram.
$\GLup$ and $\GRup$ denote
the weak vertex structure, given in NDR and in HV respectively by:
\be
\begin{array}{lr}
{\GLup=\left\{\begin{array}{cc}
\Gmuup(1\!-\!\Gfive), & \,\,\,\mbox{NDR} \\
\tilde\Gmuup(1\!-\!\Gfive), & \,\,\,\mbox{HV}
\end{array}\right.} &
{\GRup=\left\{\begin{array}{cc}
\Gmuup(1\!+\!\Gfive), & \,\,\,\mbox{NDR} \\
\tilde\Gmuup(1\!+\!\Gfive), & \,\,\,\mbox{HV}
\end{array}\right.}
\end{array}
\ee
By reducing the $\gamma$-algebra in (\ref{vertexCT_1}),
 we can separate the
contributions of 4-dimensional operators from those of EO. By definition,
the contributions of EO correspond to those
terms which vanish under a suitable projection on the
four dimensional basis. There are several possible choices of the
projection operators, which in general will define different
renormalized operators,  i.e. they will give different two-loop
anomalous dimension matrices. The results  can
be easily related, by computing at one-loop the renormalized operators
obtained by different projections and we have checked
the consistency of the contribution of several two-loop diagrams obtained
with different projections.
For the sake of comparison, in the following we shall use
 the same projection as in refs.\cite{bjlw2,bw}:
\begin{itemize}
\item for $\LL\longrightarrow \,\,{\Bbb P}_{ LL}=\RRn$
\item for $\LR\longrightarrow \,\,{\Bbb P}_{ LR}=\SP$
\end{itemize}
Then, in order to project on $\LL$,
we take the following trace:
\be
{\Bbb P}_{ LL}\left(\LL\right)=
Tr \left[\Gmuup(1\!-\!\Gfive)
\Gnuup(1\!+\!\Gfive) \Gmudw(1\!-\!\Gfive)
\Gnudw(1\!+\!\Gfive)\right)]
\label{LLprojection}
\ee
while, in order to project on $\LR$, we use:
\be
{\Bbb P}_{ LR}\left(\LR\right)=
Tr \left[\Gmuup(1\!-\!\Gfive)
(1\!-\!\Gfive) \Gmudw(1\!+\!\Gfive)
(1\!+\!\Gfive) \right]
\label{LRprojection}
\ee
In eq.(\ref{LLprojection}) the sum over $\nu$ is intended in 4-dimension
in HV and in D-dimensions in NDR.
When projecting a string of $\gamma$-matrices,  the values
of the traces in (\ref{LLprojection}) and (\ref{LRprojection}) are taken as
normalization factors.
With the projectors introduced in eq.(\ref{LLprojection})
and eq.(\ref{LRprojection}),
 we obtain the following decomposition for
the one-loop vertex diagrams in (\ref{vertexCT_1}):
\bea
G^{ (1)}_{ LL,LR}&=&C^{ (1)}\frac{1}{2\epsilon}\left(
F^{ (1)}_{ LL,LR}(\epsilon)\LLR+E^{ (1)}_{ LL,LR}\right)\nn\\
G^{ (2)}_{ LL,LR}&=&C^{ (2)}\frac{1}{2\epsilon}\left(
F^{ (2)}_{ LL,LR}(\epsilon)\LLR+E^{ (2)}_{ LL,LR}\right)\\
G^{ (3)}_{ LL,LR}&=&C^{ (3)}\frac{1}{2\epsilon}\left(
F^{ (3)}_{ LL,LR}(\epsilon)\LLR+E^{ (3)}_{ LL,LR}\right)\nn
\label{vertexCT_2}
\eea
where the $\epsilon$-dependent coefficients are given,
up to $O(\epsilon)$, by:
\bea
F^{ (1)}_{ LL}(\epsilon)=F^{ (3)}_{ LL}(\epsilon)&=&
\left\{\begin{array}{cc}
4(1-2 \epsilon) & \,\,\,\,\,\mbox{NDR} \\
4 & \,\,\,\,\,\mbox{HV}
\end{array}\right. \\
F^{ (2)}_{ LL}(\epsilon)&=&\left\{\begin{array}{cc}
-4(4-\epsilon) & \,\,\,\,\,\mbox{NDR} \\
-4(4-\epsilon)  & \,\,\,\,\,\mbox{HV}
\end{array}\right.
\label{LLcoeff}
\eea
and:
\bea
F^{ (1)}_{ LR}(\epsilon)&=&
\left\{\begin{array}{cc}
4(1-2 \epsilon) & \,\,\,\,\,\mbox{NDR} \\
4 & \,\,\,\,\,\mbox{HV}
\end{array}\right. \\
F^{ (2)}_{ LR}(\epsilon)&=&\left\{\begin{array}{cc}
-4(1+\epsilon) & \,\,\,\,\,\mbox{NDR} \\
-4(1-\epsilon)  & \,\,\,\,\,\mbox{HV}
\end{array}\right. \\
F^{ (3)}_{ LR}(\epsilon)&=&\left\{\begin{array}{cc}
16(1-\epsilon) & \,\,\,\,\,\mbox{NDR} \\
16 & \,\,\,\,\,\mbox{HV}
\end{array}\right.
\label{LRcoeff}
\eea
$E^{ (i)}_{ LL,LR}$ are the contributions due to the
EO.
\par From eq.(\ref{vertexCT_2}), we obtain the counter-terms to left-left
and left-right operators, including those proportional to EO:
\bea
\hat G^{ (1)}_{ LL}&=&C^{ (1)}\frac{1}{2\epsilon}\left(4\,
\LL+E^{ (1)}_{ LL}\right)\nn\\
\hat G^{ (2)}_{ LL}&=&C^{ (2)}\frac{1}{2\epsilon}\left(-16 \,
\LL+E^{ (2)}_{ LL}\right) \nn \\
\hat G^{ (3)}_{ LL}&=&C^{ (3)}\frac{1}{2\epsilon}\left(4 \,
\LL+E^{ (3)}_{ LL}\right)
\label{vvct}
\eea
and similarly for the LR case.
 By subtracting the counter-terms in eq.(\ref{vvct}) from the
result of the calculation of the diagrams in figs.3a-c, we thus obtain
the renormalized operator matrix elements expressed in terms of
the tree-level matrix elements  and of the matrix $\hat r$ ($\hat s$),
see eq.(\ref{coe2}). The pole and finite terms  coming from the
calculation of  diagrams
$V_{1,3}$ are reported in table \ref{t1loop} both in
HV and NDR, for the insertion of
a $\LL$ and $\LR$ operator. Only the terms proportional to the
original operators are shown. \par Notice that the insertion of an
operator $\gamma_L^{\mu}\otimes \gamma_{\mu}$ does not produce
``effervescent" operators when we compute the vertex
renormalization of the vector current $\gamma_{\mu}$, diagram $V_1$,
 since $\gamma_5$ is not involved.
As a consequence, in HV, using the $\overline{MS}$
subtraction procedure, the finite term of the operator $\gamma_L^{\mu}
\otimes \gamma_{\mu}$
is different from the finite term of the operator
$1/2 \Bigl( \gamma_L^{\mu}\otimes\gamma_{\mu L} +
\gamma_L^{\mu}\otimes \gamma_{\mu R}  \Bigr) $.
In particular for the vertex renormalization of $\gamma_{\mu}$
one obtains:
\be \frac{\alphas}{4 \pi} \Bigl(
\frac {1}{\epsilon} + \frac{1}{2} \Bigr)
\times  \gamma_L^{\mu}
\otimes \gamma_{\mu} \label{bht} \ee
instead of:
\be \frac{\alphas}{4 \pi} \Bigl(
\frac {1}{\epsilon} + \frac{5}{2}
\Bigr)  \times \frac{1}{2} \Bigl( \gamma_L^{\mu}\otimes\gamma_{\mu L} +
\gamma_L^{\mu}\otimes \gamma_{\mu R} \Bigr) \label{bht1} \ee
as can be read in table \ref{t1loop}.
This difference has important consequences as will be discussed
in the following.
\begin{table}
\begin{center}
\begin{tabular}{|c|c|ccc|ccc|}\hline\hline
\multicolumn{1}{|c}{diagram} & \multicolumn{1}{|c}{M}
&\multicolumn{3}{|c|}{$\LL$} &
\multicolumn{3}{c|}{$\LR$}
\\
\hline
 & & $(1/\epsilon)$ &
$O(1)_{HV}$ &
$O(1)_{NDR}$
& \phantom{m}$(1/\epsilon)$ &
$O(1)_{HV}$ &
$O(1)_{NDR}$\\ \hline
$V_1$ & 2&1 &5/2 &1/2 &1& 5/2 &1/2\\
$V_2$ &2& -4& -9 & -9 & -1 & -3/2&-7/2 \\
$V_3$ &2& 1 & 5/2&1/2 & 4 & 10 & 6 \\
$P_1$ & 1&-4/3 & -20/9 & -8/9 & - & - & -\\
$F_1$ & 1&-4/3 & -20/9 & -20/9 & -4/3 & -20/9 & -20/9\\
\hline\hline
\end{tabular}
\caption[]{Singular and finite terms for diagrams in
figs.3a-c and 9, with a $\LL$ or a $\LR$ Dirac structure.
The multiplicity of the diagrams  is also reported in the table.
 Colour-charge factors and a
the factor $\alphas / 4 \pi$ ($\alphae / 4 \pi$)
are omitted.}
\label{t1loop}
\end{center}
\end{table}

\subsection{ One-Gluon/Photon Penguin Diagrams at One Loop.}
\label{sec:an5}
The one-loop penguin diagrams are much simpler, due to the
absence of EO  at the one-loop level.
The computation of the diagrams in fig.9 gives, both in the HV and in
the NDR schemes, the following Dirac structure
(we omit the colour-charge factors):
\be
\left(\mbox{Penguin}\right)=-
\frac{4}{3}\frac{F_{ P}^{HV,NDR}(\epsilon)}{\epsilon}
\left(q^{ 2}\GLup-
q^{\mu}q_{ L}\!\!\!\!\!/\,\,\right)
\label{penguinCT}
\ee
where $q^{\mu}$ is the momentum of the gluon/photon
(see fig.9) and
$\GLup$ is  4-dimensional or  D-dimensional  in the HV or
in the NDR case respectively. We find:
\bea F_{ P}^{HV}(\epsilon)&=&1+\frac{5}{3}\epsilon \nn \\
 F_{ P}^{NDR}(\epsilon)&=&1+ \frac{2}{3} \epsilon \label{fpfp} \eea
We conclude that both in HV and NDR the counter-term
 has the following form:
\be
\hat G_{1g} =
\frac{-4}{3}\frac{1}{\epsilon}\left(q^{ 2}
\Gamma^{\mu}_L -q^{\mu}q\!\!\!/\,\right)
\label{pct}
\ee

\subsection{ Two-Gluon or One-Gluon+One-Photon Penguin One Loop
Diagrams}
\label{sec:an6}
The presence of two gluon contributions is a typical feature of the two-loop
calculation. The two-gluon (one gluon+one photon) diagrams, fig.12c,
only enter as counter-terms at the two-loop level, see for example
fig.15. These counter-terms
exist only if, in any two-loop penguin-like diagram,
 we subtract completely the divergent part of the internal sub-diagram,
without making use of the equations of motion. This is also true
for the longitudinal component $\sim q^{\mu}q\!\!\!/$ of the
one gluon (photon) counter-term, which vanishes by  the equations of motion.
However one may choose to subtract only counter-terms which
do not vanish by the equations of motion. The contribution of
single diagrams will be modified, but the final result will be the same.
Thus, for example, consider the diagrams $P_2$ and $P_3$.
They  can be computed
with or without the two gluon counter-terms, $\hat G^{a}_{2g}$ and
$\hat G^{b}_{2g}$ respectively,
whose sum is equal to zero, fig.16.
The abelian part, i.e. the sum of the two "complete" diagrams,
remains the same\footnote{ That is their sum is the same with or
without the subtraction of $\hat G^{a,b}_{2g}$.}
 as schematically indicated in fig.16, even though
each of them is modified.
Something similar happens to the longitudinal contribution of the
penguin counter-term, eq.(\ref{pct}), for  diagrams $P_{14}$ and $P_{15}$. If
we subtract all the terms $\sim 1/\epsilon$ in the penguin
internal sub-diagram, we are also subtracting the longitudinal
term $q^{\mu}q\!\!\!/$.  We may however make use of the equations
of motion and subtract only the term $\sim q^2 \gamma^{\mu}_L$.
 The sum of the two diagrams (corresponding
to the abelian case) remains the same
because of the cancellation of the contributions of
 the counter-terms due to the longitudinal
components, fig.17. The same happens in the non-abelian
case, but in a more complicated way which involves the difference of
the two diagrams and
also non-abelian diagrams. We have explicitely checked
that, if we subtract only those counter-terms which do not
vanish by  the equations of motion, we obtain the same anomalous
dimension matrix. However we  will give the results by subtracting all
the pole parts of the internal sub-diagrams because in this way
eq.(\ref{gdiff})
remains valid diagram by diagram, see below. \par
The cancellation of the sum of the counter-terms
proportional to the longitudinal contribution in eq.(\ref{pct})
and of the two-gluon (one gluon-one photon) counter-terms
corresponds to the substitution\cite{witten}:
\be \bar \psi_1 \gamma^{\mu}_L D^{\nu} F_{\mu,\nu} \psi_2
\rightarrow  (\bar \psi_1 \gamma^{\mu}_L \psi_2 ) \sum_{q} \bar q
\gamma_{\mu} q \label{eom} \ee
i.e. to the subtraction done using only four-fermion operators.
Notice however that minimal subtraction of the pole term proportional
to the operator $\bar \psi_1 \gamma^{\mu}_L D^{\nu} F_{\mu,\nu} \psi_2
$ in HV, is non-minimal in the basis where we use the $\LL$
and $\LR$ operators. This happens because, as explained above,
the finite term of the renormalized operator $\gamma_L^{\mu}\otimes
 \gamma_{\mu}$
is different from the finite term of the operator
$ \frac{1}{2} \Bigl( \gamma_L^{\mu}\otimes\gamma_{\mu L}+
\gamma_L^{\mu} \otimes\gamma_{\mu R} \Bigr) $. It is this last operator
that we have indeed to subract for consistency with the current-current
diagrams. The difference between $\gamma_L^{\mu}\otimes
 \gamma_{\mu}$ and $ \frac{1}{2} \Bigl( \gamma_L^{\mu}\otimes\gamma_{\mu L}+
\gamma_L^{\mu} \otimes\gamma_{\mu R} \Bigr) $ is important only for
the counter-terms of diagrams $P_{16}$ and $F_{16}$. Since we have to apply
a non-minimal subtraction to the lower vertex,
the contribution  of $P_{16}$ and $F_{16}$ is non-zero, but it is given
by one half of the pole of the one-loop penguin
diagram times the difference of the finite terms of the two operators
$\gamma_L^{\mu}\otimes
 \gamma_{\mu}$ and $ \frac{1}{2} \Bigl( \gamma_L^{\mu}\otimes\gamma_{\mu L}+
\gamma_L^{\mu} \otimes\gamma_{\mu R} \Bigr) $:
\be \frac {1}{2} \times\frac{-4}{3\epsilon} \times \left(-2\right)
= \frac{4}{3}  \frac{1}{\epsilon} \ee
in agreement with eq.(\ref{2loopADM}). This point was overlooked in
ref.\cite{bjlw1}.
\par The results in Appendix  have been
obtained by subtracting the pole term of the internal divergent
sub-diagrams, without making use of the equations of motion.
With this choice, the relation between the anomalous dimension in NDR
and HV remains valid diagram by diagram, eq.(\ref{gdiff}).
We have checked that our results satisfy eq.(\ref{gdiff})
 diagram by diagram, as the reader can verify by himself
using the results from tables \ref{t1loop}
and \ref{vvll}-\ref{flr}.
Below we will discuss how this works in three specific
examples. \par The fact that the
sum of all the counter-terms due to operators which vanish by the
equations of motion cancel, makes the relation (\ref{gdiff}) valid also for the
complete matrix. \par For completeness we report also the
explicit form of the two-gluon counter-term.
The pole part, obtained from the first diagram in figs.12c, is given
by (colour-charge factors omitted):
\be
\frac{4}{3}\frac{F_{ P}^{HV,NDR}(\epsilon)}{\epsilon} Q^{\mu\nu} \nn \ee
\be Q^{\mu\nu}=
\left(g^{\mu\nu}(k_{ 1}\!\!\!\!/-k_{
2}\!\!\!\!/)_{ L}-(2k_{ 1}+k_{ 2})^{\nu}\Gmulup +(k_{ 1}+
2k_{ 2})^{\mu}\Gnuup_{ L}\right)
\label{twogluonCT}
\ee
and the finite part, i.e. $F_{ P}^{HV,NDR}(\epsilon)$
 is (as expected) equal to the finite part
obtained from the diagram of fig.12b, cf. eq.(\ref{fpfp}).
We thus find that the counter-term
 is in this case:
\be \hat G^a_{2g} = - \hat  G^b_{2g} =
\frac{4}{3}\frac{1}{\epsilon}Q^{\mu\nu}
\label{p2ct}
\ee
A diagramatic representation of the one-loop diagrams and
the corresponding counter-terms is given in figs.12a-c.

\subsection{Two-Loop Current-Current Diagrams.}
\label{sec:2lcc}
After the detailed study of the one loop diagrams, we are
ready to show the construction of a ``complete" two-loop
diagram, including all necessary counter-terms. We will consider
as an example the diagram of fig.10 denoted as $V_{17}$, where we
insert a $\LL$ vertex. We denote by $D^{ (17)}_{ LL}$ the
double and single pole contribution from this diagram. We
then substitute to the internal loop, including the string
of Dirac matrices, the suitable
counter-term, $\hat G_{LL}^{(1)}$. We denote by
$C^{ (17)}_{C, LL}$ the
double and single pole contribution from this counter-term.
 We have finally to correct for
the mixing between 4-dimensional operators and EO which occurs
at one loop. This corresponds to the last term in eq.(\ref{2loopADM}),
the formula which gives the two-loop anomalous dimension\footnote{
The last term in eq.(\ref{2loopADM}) receives a contribution also
from non "effervescent" operators in the case of a non minimal
subtraction. In this respect there is not much difference between EO and
finite subtractions of 4-dimensional operators.}.
To obtain this term we substitute to the internal loop
the combination :
 \bea \hat E_{LL}^{(17)} &=&
- \frac{1}{2}  C^{(1)} \left( \frac{
\Gnuup\Group\GLup\Grodw\Gnudw\otimes\GLLdw
}{
4 \epsilon } \frac{F^{(1)}_{LL}(0)}{F^{(1)}_{LL}(\epsilon)}
-\frac{\GLup}{\epsilon} \right) \nn \\
&\sim&  -C^{(1)} \times \frac {1}{8 \epsilon} E^{(1)}_{LL}  \label{fri}\eea
The above equality is valid
up to terms which do not contribute to the two-loop anomalous
dimension because are of higher order in $\epsilon$, cf. eq.(\ref{vvct}).
 We denote by
 $E^{ (17)}_{C, LL}$ the  contribution from
this insertion. The contribution proportional to the first coefficient
of the beta function $\beta_0$ in eq.(\ref{2loopADM}) is absent
for the diagram $V_{17}$. It will only give a contribution for
those diagrams, which contain an internal loop  corresponding
to the renormalization of $\alphas$, as for example $V_{12}$.
The term $\sim \beta_0$ is authomatically taken into account
by subtracting the counter-term corresponding to the renormalization
of the strong vertex. \par The result for the ``complete" diagram
is thus given by the sum of the three different contributions,
see fig.13:
\be
\bar D^{(17)}_{LL}=
D^{(17)}_{LL}-C^{ (17)}_{C,LL}+ E^{ (17)}_{C,LL}
\label{VCTsubtraction}
\ee
This ends our discussion of current-current  counter-terms.
In tables \ref{vvll}
and \ref{vvlr}
we report in units of $(\alphas/4 \pi)^2$,
($\alphas \alphae / 16 \pi^2$), the double and single pole contribution
of all the current-current diagrams and
of the corresponding counter-terms, specifying the
contribution of  EO counter-terms whenever present. \par From table
\ref{vvll} we notice that, in HV,
 diagrams which can be changed
 one into the other by Fierz rearrangement, like
$V_{17}$ and $V_{20}$ for example,
give exactly the same contribution
to the anomalous dimension  (double and single poles). This is
however true only for the ``complete" diagrams,
$\bar D^{(17)}_{LL}$ and $\bar D^{(20)}_{LL}$, while there is no
relation between the corresponding bare diagram contributions,
$D^{(17)}_{LL}$ and  $D^{(20)}_{LL}$.
This happens because the HV regularization scheme preserves the
Fierz properties of the four dimensional basis of the renormalized
operators defined via $\overline{MS}$ subtraction\footnote{This is
strictly true with the projection operators  (\ref{LLprojection}) and
(\ref{LRprojection}).
Also in HV  one can choose projection operators such that
Fierz rearrangement relations  are not satisfied at two loops.}
The renormalized operators
are however obtained only in combination with all the possible
counter-terms. On the other hand, $\overline{MS}$  NDR does not
respect the Fierz properties of the renormalized operators, as
it can be seen from the results reported in
tables \ref{vvll}-\ref{flr}.
It would be possible however to obtain a two-loop anomalous dimension
which respects the Fierz symmetry by a suitable finite one-loop
redefinition of the renormalized operators\cite{acmp,bw}.
Since Fierz rearrangement is not valid in NDR, $\LR$ and
$-2 (1-\gamma_5)\otimes(1+\gamma_5)$ Fierz  rearranged operators
renormalize differently.
Notice that in the case of
$-2 (1-\gamma_5)\otimes(1+\gamma_5)$ operators,
 diagrams denoted by $P_i$ in fig.11 do not
vanish. We have also computed the renormalization of the effective
Hamiltonian in the Fierz rearranged basis and verified that
we obtain consistent results, i.e. that  the anomalous
dimension matrices in the basis (\ref{epsilonprime_basis}) and in the
rearranged one satisfy eq.(\ref{gdiff}).
\par As already anticipated, the relation which connects the
two-loop anomalous dimension in two different regularization schemes
(\ref{gdiff}) holds diagram by diagram. This means that in
eq.(\ref{gdiff}) (or in eq.(\ref{gamma1})), we have to interpret
the two-loop anomalous dimension $\hat \gamma^{(1)}$ (or $\Z_1^{(2)}$)
as the contribution of the particular diagram under consideration.
We have also to interpret
 the one-loop anomalous dimension $\hat \gamma^{(0)}$ (or $\Z_1^{(1)}$)
and one-loop matrix $\hat r$ (or  $\Z_0^{(1)}$) as due to all
the possible diagrams obtained by eliminating one of the gluon
propagators in the two-loop diagram. \par
We illustrate  further this point
by discussing again the current-current diagram $V_{17}$ of fig.10.
 Eq.(\ref{gdiff}), for a single diagram, has to be interpreted as
follows. The commutator $\Bigl[ \Delta \hat r , \gamma^{(0)}_s
\Bigr] $ corresponds to the difference of the finite parts of the
internal diagram $\Delta r_{V_1}$, corresponding to diagram $V_1$, times
the pole term of the external diagram $p_{V_2}$ minus the difference
of the finite
parts of the external diagram $\Delta r_{V_2}$ times $p_{V_1}$.
Omitting
colour-charge factors, in the case at hand we have:
\bea \Delta p^{(1)}= p^{(1)}_{NDR}-p^{(1)}_{HV}&=&\frac{1}{2} \left(
 \Delta r_{V_1} p^{(0)}_{V_2} -  p^{(0)}_{V_1} \Delta r_{V_2} \right)
\rightarrow
\nn \\ \frac{7}{2} - (-\frac{1}{2}) &=&\frac{1}{2}
\left( (-2) \times (-4) - 1 \times 0 \right) \eea
which coincides with the difference of the $1/ \epsilon$ term
reported in table \ref{vvll} for $V_{17}$.  \par
In general the appropriate formula is:
\be \Delta p^{(1)} =
p^{(1)}_{NDR}-p^{(1)}_{HV}= \frac{1}{2} \sum_{a,b} \left[
\Delta r_a \cdot p^{(0)}_b -p^{(0)}_a  \cdot \Delta r_b  \right]
 \label{test}\ee
$a$ and $b$ indicate all the possible sub-diagrams which appear
in the two-loop diagram.
$p^{(1)}_{NDR,HV}$, which is the single pole of the two-loop diagram,
 has a subscript because it depends on the regularization.
$p^{(0)}_a$ is the one-loop pole term for the diagram $a$.
$\Delta r_a$ is
the difference of the finite terms between NDR  and HV for the sub-diagram $a$.
If $p^{(0)}_a$ includes the poles corresponding to the renormalization
of the strong coupling constant, eq.(\ref{test}) incorporates authomatically
the last term of eq.(\ref{gdiff}).
We have verified diagram by diagram that all our
results satisfy eq.(\ref{test}).

\subsection{Two-Loop Penguin Diagrams.}
\label{sec:2lp}
 We will consider,
as an example of the two-loop penguin diagrams,
the diagram of fig.11 denoted as $P_3$, where we
insert a $\LL$ vertex. We denote by $P^{(3)}_{LL}$ the
double and single pole contribution from this diagram. We
then substitute to the internal loop, including the string
of Dirac matrices, the suitable
counter-term, $\hat G_{LL}^{(2)}$.
 In this case however we have also to subtract the
two-gluon counter-term since  the first of the two diagrams in
fig.12c is also contained in $P_3$ as a divergent sub-diagram.
We denote by
$C^{ (3)}_{P, LL}$ the
double and single pole contribution from all the  counter-terms.
 We have finally to correct for
the mixing between 4-dimensional operators and EO which occurs
at one loop. This corresponds to the last term in eq.(\ref{2loopADM}),
the formula which gives the two-loop anomalous dimension.
As was the case for $V_{17}$,
to obtain this term, we substitute to the internal loop
the combination:

 \bea \hat E_{LL}^{(3)} &=&
- \frac{1}{2}  C^{(2)} \left( \frac{
\GLup\Gnuup\Group\otimes\GLLdw
\Gnudw\Grodw}{
4 \epsilon } \frac{F^{(2)}_{LL}(0)}{F^{(2)}_{LL}(\epsilon)}
-\frac{(-4 \GLup)}{\epsilon} \right) \nn \\
&\sim & -C^{(2)} \times \frac {1}{8 \epsilon} E^{(2)}_{LL}  \label{fro} \eea
and denote by
$E^{ (3)}_{P, LL}$ the  double and single pole contribution from
this insertion. We thus obtain:
\be
\bar P^{(3)}_{LL}=
P^{(3)}_{LL}-C^{ (3)}_{ P,LL}+ E^{ (3)}_{P,LL}
\label{Psubtraction}
\ee
A diagramatic representation of the subtraction procedure
is reported in fig.14. In HV two diagrams which go one into
the other by Fierz rearrangement give the same contribution
to the anomalous dimension. All the results for the  ``complete" penguin
diagrams are given in tables
\ref{pll}-\ref{flr} for HV and  for NDR.
\par  We can check eq.(\ref{gdiff}) also in the case of $P_3$. We call
$\Delta r_{2g}$ and $p_{2g}$ the finite part and pole term of the
two-gluon diagram of fig.12c  and $\Delta r_{2gct}$ and $p_{2gct}$
the corresponding quantities for the insertion of the operator  $Q^{\mu\nu}$ of
 eq.(\ref{twogluonCT}), see fig.15.
One then obtains ($p_{2g}=-4/3$, $p_{2gct}=1$, $\Delta r_{2g}=4/3$
and $\Delta r_{2gct}=0$):
\be \Delta p^{(1)} = \frac{1}{2} \left(\Delta r_{V_2} p^{(0)}_{P_1} -
p^{(0)}_{V_2} \Delta r_{P_1} + \Delta r_{2g} p_{2gct} - p_{2g} \Delta r_{2gct}
\right) = \frac{10}{3}  \label{pippo} \ee
in agreement with the results of table \ref{pll}\footnote{When
we take into account the multiplicity.}.
\par We notice that our results for  the diagrams $P_8$ and
$F_8$ in HV satisfy eq.(\ref{gdiff}),  as can be easily verified.
For these diagrams, in HV, the authors of ref.\cite{bjlw1} found a
different result
which  fails to satisfy eq.(\ref{gdiff}).
\subsection{Terms Which Vanish by Equations of Motion
and Gauge Invariance}
\label{sec:eom}
The results reported in tables \ref{vvll}-\ref{flr}
 have been obtained
by subtracting the pole term, including the EO,
eqs.(\ref{vvct}), for all the divergent sub-diagrams.
In doing so, we have subtracted also operators which vanish by
the equations of motion or non-gauge invariant operators.
As already discussed before, we could have subtracted only
four-fermion operators and obtained the same final result.
However contributions of single diagrams would have been different
and the check between HV and NDR, eq.(\ref{gdiff}), would have not
be valid diagram by diagram. Besides this we have of course to take
into account the mixing with ``effervescent" operators. This
mixing is responsible, besides other effects, of the non-vaninshing
of diagrams $P_{16}$ and $F_{16}$ in HV.
\par  The presence of operators which vanish by  the equations of motion
and non-gauge  invariant
operators is signaled by the appearence of various tensor
products in the calculation of the different two-loop
Feynman diagrams (we follow the notation of ref.\cite{bjlw2}):
\bea T_1 &=& \gamma_{\mu}^L \otimes \gamma^{\mu} \nn \\
T_2 &=& = q\!\!\!/ \left( 1 - \gamma_5 \right) \otimes
q\!\!\!/ \frac {1}{q^2} \nn \\
T_3 &=& T_1 \times \frac{p \cdot r} {q^2} \\
T_4 &=& \Bigl( p\!\!\!/ \left( 1 - \gamma_5 \right) \otimes
r\!\!\!/ + r\!\!\!/ \left( 1 - \gamma_5 \right) \otimes
p\!\!\!/ \Bigr) \frac{1}{q^2} \nn \\
T_5 &=& S_{\mu \sigma \nu} p^{\mu} r^{\nu} \left( 1 -\gamma_5
\right) \otimes \gamma^{\sigma} \frac {1}{q^2} \nn \eea
where $-p$ is the ingoing momentum, $r$ the outgoing momentum,
$q=p+r$ and $S_{\mu \sigma \nu}= \gamma_{\mu}
\gamma_{\sigma} \gamma_{\nu}- \gamma_{\nu} \gamma_{\sigma}
\gamma_{\mu}$. \par
$T_2$ and $T_4$ vanish by using the equations of motion:
\be \bar u(r) r\!\!\!/ =p\!\!\!/ u(p)=0 \ee and
\be T_3 \rightarrow \frac {T_1}{2} \,\,\,\,\,\,\,\,\,\,\,\,
T_5 \rightarrow  - T_1 \ee
\par Non-gauge invariant operators can be eliminated by using
the background field Feynman gauge\cite{abbott} for the non-abelian
penguin diagrams, for example $P_4$ and $P_6$ in fig.11.
This allows simple checks of the gauge invariance
of the final result,
 because non-gauge invariant operators cancel when combining
together sub-sets of two-loop penguin diagrams. \par
We give an explicit example of this cancellation by considering
diagrams $P_2$, $P_3$ and $P_4$ in HV.
The results for these diagrams, using the background field
gauge, putting $p^2=r^2=0$ and reporting only double and
single poles are:
\bea \bar P_2^{\mu} &=& \left( \frac{1}{2 \epsilon^2}
-\frac{59}{36 \epsilon} \right) q^2 \gamma^{\mu}_L +
\left( -\frac{2}{3 \epsilon^2} +\frac{1}{9 \epsilon} \right)
q\!\!\!/_L q^{\mu} \nn \\ &+&
\left(- \frac{1}{3 \epsilon^2} +\frac{29}{18 \epsilon} \right)
q\!\!\!/_L p^{\mu}
+\left( -\frac{1}{3 \epsilon^2}
+\frac{5}{18 \epsilon} \right) p\!\!\!/_L q^{\mu}
\nn \\
&+& \left( \frac{4}{9 \epsilon^2}
-\frac{16}{27\epsilon} \right) p\!\!\!/_L p^{\mu}
+ \left( \frac{1}{\epsilon^2}+ \frac{1}{2 \epsilon} \right) \, i \,
\epsilon^{\mu \nu \rho \sigma}q_{\nu}p_{\rho}\gamma_{\sigma}^L \nn \\ &+&
 ( p \leftrightarrow r)  \nn
\eea
\bea \bar P_3^{\mu} &=& \left( -\frac{5}{2 \epsilon^2}
+\frac{59}{36 \epsilon} \right) q^2 \gamma^{\mu}_L +
\left( \frac{8}{3 \epsilon^2} -\frac{10}{9 \epsilon} \right)
q\!\!\!/_L q^{\mu} \nn \\ &+&
\left( \frac{1}{3 \epsilon^2} +\frac{7}{18 \epsilon} \right)
q\!\!\!/_L p^{\mu}
+\left( \frac{1}{3 \epsilon^2}
-\frac{5}{18 \epsilon} \right) p\!\!\!/_L q^{\mu}
 \nn \\
&+& \left( -\frac{4}{9 \epsilon^2}
+\frac{16}{27 \epsilon} \right) p\!\!\!/_L p^{\mu}
+ \left( -\frac{1}{\epsilon^2}+ \frac{3}{2 \epsilon} \right)\, i \,
\epsilon^{\mu \nu \rho \sigma}q_{\nu}p_{\rho}\gamma_{\sigma}^L\nn
\nn \\ &+&  ( p \leftrightarrow r)  \nn
\eea
\bea \bar P_4^{\mu} &=& \left( \frac{8}{9 \epsilon^2}
+\frac{13}{27 \epsilon} \right) q^2 \gamma^{\mu}_L +
\left( -\frac{5}{9 \epsilon^2} +\frac{25}{54 \epsilon} \right)
q\!\!\!/_L q^{\mu} \nn \\ &-&
\left( \frac{4}{9 \epsilon} \right)
q\!\!\!/_L p^{\mu}
-\left(
\frac{2}{9 \epsilon} \right) p\!\!\!/_L q^{\mu} \nn \\
&+& \left( \frac{2}{9 \epsilon^2}
+\frac{13}{27 \epsilon} \right) p\!\!\!/_L p^{\mu}
+ \left( -\frac{2}{ 3 \epsilon^2}+ \frac{13}{9 \epsilon} \right)\, i \,
\epsilon^{\mu \nu \rho \sigma}q_{\nu}p_{\rho}\gamma_{\sigma}^L \nn
\\ &+&  ( p \leftrightarrow r)  \nn
\eea
The explicit results are given for  a gluon propagator stemming from
the upper incoming quark leg, as shown in fig.11. $p$ is
replaced by $r$ (and the term proportional to
$\epsilon_{\mu \nu \rho \sigma}$ changes sign)
when the gluon is attached to the outgoing quark leg.
Notice that $q_{\mu} \bar P_2^{\mu}$ is different from zero and
similarly for $\bar P_3^{\mu}$ and $\bar P_4^{\mu}$,
indicating the presence
of non-gauge invariant operators\footnote{We use the on-shell identity
$p \cdot q = q^2/2$}.
However:
\bea q_{\mu} \left( \bar P_2^{\mu}+\bar P_3^{\mu} \right) =0 \nn \\
q_{\mu} \left(\bar P_2^{\mu}-\bar P_3^{\mu}  + 2 \bar P_4^{\mu}
\right) =0 \nn \eea
if we use the backgroud field gauge for the non-abelian diagram.
The sum of the above diagrams eliminates the non-gauge invariant
operators. By writing the term $\sim \epsilon_{\mu \nu \rho \sigma}$
as a term proportional to $q^2 \gamma^{\mu}_L - q\!\!\!/_L q^{\mu}$
plus terms which vanish by the equations of motion  and by eliminating
for the same reason
all terms proportional to $p\!\!\!/$ or $r\!\!\!/$, we
arrive to the results reported in the tables. \vskip 0.5 cm \par
We conclude this section by summarizing some checks done to
verify the correctness of our calculation.
\par 1) We have verified that the $1/\epsilon^2$ contribution of the
counter-term is twice the corresponding term of the original
diagram, as imposed by eq.(\ref{twoloop_ad_condition}).
\par 2) We have verified the cancellation of all the single poles
accompanied by logarithms of the external momenta, i.e.
$\sim 1/\epsilon \times ln(p^2/\mu^2),1/\epsilon \times ln(r^2/\mu^2)$
or $1/\epsilon \times ln(q^2/\mu^2)$. For penguin diagrams, this
cancellation does not follow authomatically from 1).
\par 3) We have verified the cancellation of all non-gauge invariant
terms in the background Feynman gauge and repeated the calculation
in the standard Feynman gauge with the same final result.
\par 4) We have verified that we get the same result
 by subtracting all the pole terms
in the internal loops, i.e. by subtracting also operators which
vanish by the equations of motion, or by subtracting only
those counter-terms which are proportional to four-fermion operators.
\par 5) We have verified that eq.(\ref{gdiff}), which relates
the NDR and HV schemes, is valid diagram
by diagram. We have also verified that  eq.(\ref{gdiff}) is valid for
the matrices $\hat \gamma_s^{(1)}$ and $\hat \gamma_e^{(1)}$.
 \par In doing so we have found a difference in the result
of diagram $P_8$ ($F_8$)
between our calculation and the calculation
reported in ref.\cite{bjlw1}, as can be read from tables
\ref{pll}-\ref{flr}.
Our results for these diagrams, $P_8$  and $F_8$,
 satisfy eq.(\ref{gdiff}), while
the results of ref.\cite{bjlw1} do not. Moreover the authors
of ref.\cite{bjlw1} overlooked the different renormalization of
the operator  $\gamma_L^{\mu}\otimes \gamma_{\mu}$
  with respect to the renormalization of
$1/2 \Bigl( \gamma_L^{\mu}\otimes\gamma_{\mu L} +
\gamma_L^{\mu}\otimes \gamma_{\mu R} \Bigr) $ and consequently
do not consider the diagrams that we call $P_{16}$ and
$F_{16}$. These differences compensate in the calculation of
$\hat \gamma^{(1)}_s$. It is however easy to show that, by using for the
diagrams  the results of ref.\cite{bjlw1}, one cannot satisfy
the relation (\ref{gdiff}) for $\hat \gamma^{(1)}_e$.
As a further check we also computed the anomalous dimension matrix
at $O(\alphae^2)$, since in this case  different diagrams enter
with different weights with respect to $\hat \gamma^{(1)}_s$
and $\hat \gamma^{(1)}_e$.
We found again that only our results satisfy (\ref{gdiff}),
contrary to those of ref.\cite{bjlw1}. This gives us further
confidence on the results reported here.
\par We finally notice that in ref.\cite{bjl}, the authors have
only computed the anomalous dimension matrix in NDR. They get
the matrix in HV using eq.(\ref{gdiff}). Since we agree in NDR,
 the final result of ref.\cite{bjl} is correct, in spite of
the difference found for the  two diagrams in HV.
\subsection{Quark Self-Energy Diagrams}
\label{sec:qse}
The calculation of the quark self energy diagrams at one and two
loops is straightforward and has already been done by several
authors, see for example \cite{bw}. We only report in table \ref{1and2loopSE}
of the Appendix
the results of the different diagrams shown in figs.4, 6 and 18.
\pagebreak
\section{The anomalous dimension matrices}
\label{sec:gelettro}

In this section we collect the results of the calculation of the anomalous
dimension matrices at one- and two-loop level both in the HV and NDR
renormalization
schemes and the matrices
$\Delta \hat r$ and $\Delta \hat s$, relevant for the comparison of the results
in the two schemes.

\subsection{One-loop results}
\label{sec:oneloopres}

The one-loop anomalous dimension matrices \cite{gilman,lus} and $\Delta \hat r$
are presented here. They can be computed from the pole and finite parts
of the one-loop diagrams in figs.3a-c and 9,
given in table \ref{t1loop}, and from the diagrams of figs.4 and 6a,
table \ref{1and2loopSE}. We also report  the $\delta \hat r$ and
$\delta \hat s$ matrices introduced at the end of sec.\ref{sec:gefo}
to take into account the bottom threshold.

The one-loop $O(\alphas)$ matrix is given by

\bea
& &\hat\gamma^{\sss (0)}_{\sss s}= \\
& & \nn\\
\!& &\!\left(
\begin{array}{c c c c c c c c c c}
-\frac{6}{N} & 6 & 0 & 0 & 0 & 0 & 0 & 0 & 0 & 0\\
6 & -\frac{6}{N} & -\frac{2}{3N} & \frac{2}{3} & -\frac{2}{3N} &
\frac{2}{3} & 0 & 0 & 0 & 0 \\
0 & 0 & -\frac{22}{3N} & \frac{22}{3} & -\frac{4}{3N} &
\frac{4}{3} & 0 & 0 & 0 & 0 \\
0 & 0 & \frac{18N-2f}{3N} & \frac{-18+2Nf}{3N} & -\frac{2}{3N}f &
\frac{2}{3}f & 0 & 0 & 0 & 0\\
0 & 0 & 0 & 0 & \frac{6}{N} & -6 & 0 & 0 & 0 & 0\\
0 & 0 & -\frac{2}{3N}f & \frac{2}{3}f & -\frac{2}{3N}f &
\frac{18-18N^2+2Nf}{3N} & 0 & 0 & 0 & 0\\
0 & 0 & 0 & 0 & 0 & 0 & \frac{6}{N} & -6 & 0 & 0\\
0 & 0 & -\frac{2u-d}{3N} & \frac{2u-d}{3} & -\frac{2u-d}{3N} &
\frac{2u-d}{3} & 0 & \frac{6-6N^2}{N} & 0 & 0\\
0 & 0 & \frac{2}{3N} & -\frac{2}{3} & \frac{2}{3N} &
-\frac{2}{3}& 0 & 0 & -\frac{6}{N} & 6 \\
0 & 0 & -\frac{2u-d}{3N} & \frac{2u-d}{3} & -\frac{2u-d}{3N} &
\frac{2u-d}{3} & 0 & 0 & 6 & -\frac{6}{N}
\end{array}\right)\nn
\label{1loopADM_QCD}
\eea
 where $f$, $u$ and $d$ represent the number of flavours, the number
of charge $2/3$ up-like quarks and the number of charge $-1/3$ down-like
quarks respectively.

For the electro-magnetic anomalous dimension matrix one finds:
\pagebreak
\bea
& &\hat\gamma^{\sss (0)}_{\sss e}= \\
& & \nn\\
\!& &\!\left(\begin{array}{c c c c c c c c c c}
-\frac{8}{3} & 0 & 0 & 0 & 0 & 0 &\frac{16N}{27} & 0 & \frac{16N}{27} & 0 \\
0 & -\frac{8}{3} & 0 & 0 & 0 & 0 &\frac{16}{27} & 0 & \frac{16}{27} & 0 \\
0 & 0 & 0 & 0 & 0 & 0
&\frac{-16+(16u-8d)N}{27} & 0 & \frac{-88+(16u-8d)N}{27} & 0 \\
0 & 0 & 0 & 0 & 0 & 0 &\frac{-16N+16u-8d}{27} & 0 &
\frac{-16N+16u-8d}{27} & -\frac{8}{3} \\
0 & 0 & 0 & 0 & 0 & 0 & \frac{72+(16u-8d)N}{27} & 0 &
\frac{(16u-8d)N}{27} & 0 \\
0 & 0 & 0 & 0 & 0 & 0 & \frac{16u-8d}{27} & \frac{8}{3} &
\frac{16u-8d}{27} & 0 \\
0 & 0 & 0 & 0 & \frac{4}{3} & 0 & \frac{36+(16u+4d)N}{27} & 0 &
\frac{(16u+4d)N}{27} & 0 \\
0 & 0 & 0 & 0 & 0 & \frac{4}{3} & \frac{16u+4d}{27} & \frac{4}{3} &
\frac{16u+4d}{27} & 0 \\
0 & 0 & -\frac{4}{3} & 0 & 0 & 0 & \frac{8+(16u+4d)N}{27} & 0 &
\frac{-28+(16u+4d)N}{27} & 0 \\
0 & 0 & 0 & -\frac{4}{3} & 0 & 0 & \frac{8N+16u+4d}{27} & 0 &
\frac{8N+16u+4d}{27} & -\frac{4}{3} \nn
\end{array}\right)
\label{1loopADM_QED}
\eea

At one loop, from the finite terms in table \ref{t1loop} we also  obtain:
\bea & &\Delta \hat r= \nn\\
& & \\
\!& &\!\left(\begin{array}{c c c c c c c c c c}
\frac{4-2N^2}{N} & -2 & 0 & 0 & 0 & 0 & 0 & 0 & 0 & 0 \\
-2 & \frac{ 4-2N^2}{N} & -\frac{1}{3N} & \frac{1}{3} & -\frac{1}{3N} &
\frac{1}{3} & 0 & 0 & 0 & 0 \\
0 & 0 & \frac{10- 6N^2}{3N} & -\frac{4}{3} & -\frac{2}{3N} & \frac{2}{3} &
0 & 0 & 0 & 0 \\
0 & 0 & -2 & \frac{4-2N^2}{N} & 0 & 0 & 0 & 0 & 0 & 0 \\
0 & 0 & 0 & 0 & \frac{8-2N^2}{N} &-6 & 0 & 0 & 0 & 0 \\
0 & 0 & 0 & 0 & -4 & \frac{8-4N^2}{N} & 0 & 0 & 0 & 0 \\
0 & 0 & 0 & 0 & 0 & 0 & \frac{8-2N^2}{N} & -6 & 0 & 0 \\
0 & 0 & 0 & 0 & 0 & 0 & -4 & \frac{8-4N^2}{N} & 0 & 0 \\
0 & 0 & \frac{1}{3N} & -\frac{1}{3} & \frac{1}{3N} & -\frac{1}{3} &
0 & 0 & \frac{4-2N^2}{N} & -2 \\
0 & 0 & 0 & 0 & 0 & 0 & 0 & 0 & -2 & \frac{4-2N^2}{N} \nn
\end{array}\right)
\label{deltar_QCD}
\eea

in the strong case and
\bea
& &\Delta \hat s= \nn\\
&  & \\
\!& &\!\left(\begin{array}{c c c c c c c c c c}
-\frac{2}{9} & 0& 0 & 0 & 0 & 0 & \frac{8N}{27} & 0 & \frac{8N}{27} & 0 \\
0 & -\frac{2}{9} & 0 & 0 & 0 & 0 &\frac{8}{27} & 0 & \frac{8}{27} & 0 \\
0 & 0 & -\frac{2}{3} & 0 & 0 & 0 &-\frac{8}{27} & 0 & \frac{4}{27} & 0 \\
0 & 0 & 0 & -\frac{2}{3} & 0 & 0 &-\frac{8N}{27} & 0 & -\frac{8N}{27} &
\frac{4}{9} \\
0 & 0 & 0 & 0 & -\frac{2}{3} & 0 &\frac{20}{9} & 0 & 0 & 0 \\
0 & 0 & 0 & 0 & 0 & -\frac{2}{3} & 0 &\frac{20}{9} & 0 & 0 \\
0 & 0 & 0 & 0 & \frac{10}{9} & 0 &\frac{4}{9} & 0 & 0 & 0 \\
0 & 0 & 0 & 0 & 0 & \frac{10}{9} & 0 &\frac{4}{9} & 0 & 0 \\
0 & 0 & \frac{2}{9} & 0 & 0 & 0 & \frac{4}{27} & 0 & -\frac{8}{27} & 0 \\
0 & 0 & 0 & \frac{2}{9} & 0 & 0 & \frac{4N}{27} & 0 & \frac{4N}{27} &
-\frac{4}{9} \nn
\end{array}\right)
\label{deltar_QED}
\eea
in the electromagnetic one.

The matrices $\delta\hat r$ and $\delta\hat s$ in eq.(\ref{matm}) are given by
\bea
& &\delta \hat r= \hat r(u,d)-\hat r(u,d-1)=\nn\\
& & \\
\!& &\!\left(\begin{array}{c c c c c c c c c c}
0 & 0 & 0 & 0 & 0 & 0 & 0 & 0 & 0 & 0 \\
0 & 0 & 0 & 0 & 0 & 0 & 0 & 0 & 0 & 0 \\
0 & 0 & 0 & 0 & 0 & 0 & 0 & 0 & 0 & 0 \\
0 & 0 & \frac{5}{9N} & -\frac{5}{9} & \frac{5}{9N} & -\frac{5}{9} & 0 & 0
& 0 & 0 \\
0 & 0 & 0 & 0 & 0 & 0 & 0 & 0 & 0 & 0 \\
0 & 0 & \frac{5}{9N} & -\frac{5}{9} & \frac{5}{9N} & -\frac{5}{9} & 0 & 0
& 0 & 0 \\
0 & 0 & 0 & 0 & 0 & 0 & 0 & 0 & 0 & 0 \\
0 & 0 & -\frac{5}{18N} & \frac{5}{18} & -\frac{5}{18N} & \frac{5}{18} & 0
& 0 & 0 & 0 \\
0 & 0 & 0 & 0 & 0 & 0 & 0 & 0 & 0 & 0 \\
0 & 0 & -\frac{5}{18N} & \frac{5}{18} & -\frac{5}{18N} & \frac{5}{18} & 0
& 0 & 0 & 0
\nn
\end{array}\right)
\eea

\bea
& &\delta \hat s= \hat s(u,d)- \hat s(u,d-1)=\nn\\
& & \\
\!& &\!\left(\begin{array}{c c c c c c c c c c}
0 & 0 & 0 & 0 & 0 & 0 & 0 & 0 & 0 & 0 \\
0 & 0 & 0 & 0 & 0 & 0 & 0 & 0 & 0 & 0 \\
0 & 0 & 0 & 0 & 0 & 0 & \frac{20N}{81} & 0 & \frac{20N}{81} & 0 \\
0 & 0 & 0 & 0 & 0 & 0 & \frac{20}{81} & 0 & \frac{20}{81} & 0 \\
0 & 0 & 0 & 0 & 0 & 0 & \frac{20N}{81} & 0 & \frac{20N}{81} & 0 \\
0 & 0 & 0 & 0 & 0 & 0 & \frac{20}{81} & 0 & \frac{20}{81} & 0 \\
0 & 0 & 0 & 0 & 0 & 0 & -\frac{10N}{81} & 0 & -\frac{10N}{81} & 0 \\
0 & 0 & 0 & 0 & 0 & 0 & -\frac{10}{81} & 0 & -\frac{10}{81} & 0 \\
0 & 0 & 0 & 0 & 0 & 0 & -\frac{10N}{81} & 0 & -\frac{10N}{81} & 0 \\
0 & 0 & 0 & 0 & 0 & 0 & -\frac{10}{81} & 0 & -\frac{10}{81} & 0
\nn
\end{array}\right)
\eea
where $u$ ($d$) is the number of active u-type (d-type) quarks. These
matrices are equal in the HV and NDR schemes.

\subsection{Two-loop results in HV and NDR}
\label{sec:twoloopreshv}

All the ingredients needed for the calculation
of the two loop anomalous dimension have been given in secs.
\ref{sec:andim} and \ref{sec:an3}. Our final results for the two-loop anomalous
 dimension matrix
introduced in eq.(\ref{rge}) are
given below in the HV  and NDR regularization schemes.

\par Each  Feynman diagram can be schematically represented as
the product of three factors:
\bea ({\rm Colour-Charge} )\times ({\rm Dirac} )\times ({\rm Numerical
\,\,\, result \,\,\,
of \,\,\,loop \,\,\,integrations}) \nn  \eea
The numerical contribution of each diagram can be found in tables
\ref{vvll}-\ref{flr}, for vertex and penguin
diagrams. It
is  given in units of $\alphas/4\pi$ ($\alphae/4\pi$).
The Dirac structure can be read directly from the diagrams
with few exceptions which are explicitly reported in the tables.
The results of the easy but tedious calculation of the colour-charge factors
are not given. \par
The non-zero two-loop anomalous dimension of the weak current has been put
to zero by a suitable redefinition of the current at one-loop.
The effect of this redefinition has been already taken into account in
in the calculation of the coefficient functions.
 Without this redefinition, using the minimal $\overline{MS}$ subtraction
in  the HV scheme, we would have found $\gamma_{ J}=
(N^2-1)/2N \times 4 \beta_0$. Both the one-loop and the two-loop anomalous
dimension matrices are comprehensive of the self-energy contribution.

\newpage
\clearpage

\begin{table}
\begin{center}
\begin{tabular}{|c|c|c|}\hline
$(i,j)$ & HV & NDR \\ \hline\hline
\V$(1,1)$ & $\frac{44N^2}{3}-\frac{110}{3}-\frac{57}{2N^2}-\frac{8N}{3}f
+\frac{14}{3N}f$ & $-\frac{22}{3}-\frac{57}{2N^2}-\frac{2}{3N}f$ \\
\V$(1,2)$ & $\frac{23N}{2}+\frac{39}{N}-2f$ & $-\frac{19N}{6}+\frac{39}{N}
+\frac{2}{3}f$ \\
\V$(1,3)$ & $3N-\frac{4}{N}$ & $3N-\frac{2}{3N}$ \\
\V$(1,4)$ & $1$ & $-\frac{7}{3}$ \\
\V$(1,5)$ & $-3N+\frac{2}{N}$ & $-3N+\frac{16}{3N}$ \\
\V$(1,6)$ & $1$ & $-\frac{7}{3}$ \\[0.5ex]
\hline
\V$(2,1)$ & $\frac{23N}{2}+\frac{39}{N}-2f$ & $-\frac{19N}{6}+\frac{39}{N}
+\frac{2}{3}f$ \\
\V$(2,2)$ & $\frac{44N^2}{3}-\frac{110}{3}-\frac{57}{2N^2}-\frac{8N}{3}f
+\frac{14}{3N}f$ & $-\frac{22}{3}-\frac{57}{2N^2}-\frac{2}{3N}f$ \\
\V$(2,3)$ & $-\frac{56}{27}+\frac{86}{27N^2}$ & $-\frac{32}{27}
+\frac{86}{27N^2}$ \\
\V$(2,4)$ & $\frac{110N}{27}-\frac{140}{27N}$ &
$\frac{176N}{27}-\frac{230}{27N}$ \\
\V$(2,5)$ & $-\frac{128}{27}-\frac{58}{27N^2}$ & $-\frac{122}{27}-
\frac{94}{27N^2}$ \\
\V$(2,6)$ & $\frac{38N}{27}+\frac{148}{27N}$ &
$\frac{86N}{27}+\frac{130}{27N}$ \\[0.5ex]
\hline
\V$(3,3)$ & $\frac{44N^2}{3}-\frac{1102}{27}-\frac{1195}{54N^2}+\frac{N}{3}f
+\frac{2}{3N}f$ & $-\frac{262}{27}-\frac{1195}{54N^2}+3Nf-\frac{10}{3N}f$ \\
\V$(3,4)$ & $\frac{1061N}{54}+\frac{773}{27N}-f$ & $\frac{533N}{54}+
\frac{593}{27N}+\frac{1}{3}f$ \\
\V$(3,5)$ & $-\frac{256}{27}-\frac{116}{27N^2}-3Nf+\frac{2}{N}f$ &
$-\frac{244}{27}
-\frac{188}{27N^2}-3Nf+\frac{10}{3N}f$ \\
\V$(3,6)$ & $\frac{76N}{27}+\frac{296}{27N}+f$ &
$\frac{172N}{27}+\frac{260}{27N}
-\frac{1}{3}f$ \\[0.5ex]
\hline
\V$(4,3)$ & $\frac{35N}{2}+\frac{31}{N}-\frac{110}{27}f+\frac{86}{27N^2}f$ &
$\frac{17N}{6}+\frac{113}{3N}-\frac{2}{27}f+\frac{74}{27N^2}f$ \\
\V$(4,4)$ & $\frac{44N^2}{3}-\frac{104}{3}-\frac{57}{2N^2}+\frac{38N}{27}f-
\frac{14}{27N}f$ & $-12-\frac{57}{2N^2}+\frac{110N}{27}f-\frac{182}{27N}f$ \\
\V$(4,5)$ & $-6N+\frac{4}{N}-\frac{128}{27}f-\frac{58}{27N^2}f$ & $-6N
+\frac{32}{3N}-\frac{56}{27}f+\frac{2}{27N^2}f$ \\
\V$(4,6)$ & $2+\frac{38N}{27}f+\frac{148}{27N}f$ &
$-\frac{14}{3}+\frac{74N}{27}f
-\frac{20}{27N}f$ \\[0.5ex]
\hline
\V$(5,3)$ & $-3Nf+\frac{8}{3N}f$ & $-3Nf+\frac{20}{3N}f$ \\
\V$(5,4)$ & $\frac{1}{3}f$ & $-\frac{11}{3}f$ \\
\V$(5,5)$ & $\frac{44N^2}{3}-\frac{71}{6}+\frac{15}{2N^2}+\frac{N}{3}f$ &
$\frac{137}{6}+\frac{15}{2N^2}+3Nf-\frac{20}{3N}f$ \\
\V$(5,6)$ & $-\frac{40N}{3}+\frac{3}{N}-\frac{1}{3}f$ & $-\frac{100N}{3}
+\frac{3}{N}+\frac{11}{3}f$ \\[0.5ex]
\hline
\V$(6,3)$ & $-\frac{128}{27}f-\frac{94}{27N^2}f$ & $-\frac{56}{27}f
-\frac{178}{27N^2}f$ \\
\V$(6,4)$ & $\frac{20N}{27}f+\frac{202}{27N}f$ & $-\frac{16N}{27}f
+\frac{250}{27N}f$ \\
\V$(6,5)$ & $\frac{107N}{6}-\frac{18}{N}-\frac{74}{27}f+\frac{86}{27N^2}f$ &
$-\frac{71N}{2}-\frac{18}{N}+\frac{178}{27}f+\frac{74}{27N^2}f$ \\
\V$(6,6)$ & $-\frac{9N^2}{2}-\frac{17}{6}+\frac{15}{2N^2}+\frac{56N}{27}f-
\frac{68}{27N}f$ & $-\frac{203N^2}{6}+\frac{479}{6}+\frac{15}{2N^2}+
\frac{200N}{27}f-\frac{452}{27N}f$ \\[0.5ex]
\hline
\end{tabular}
\caption[]{Elements of the two-loop anomalous dimension matrix
$\left(\hat\gamma_{\sss s}^{\sss (1)}\right)_{ij}$.
The elements which are not reported are equal to zero. (Continue)}
\label{andim2}
\end{center}
\end{table}

\newpage
\clearpage

\addtocounter{table}{-1}
\begin{table}
\begin{center}
\begin{tabular}{|c|c|c|}\hline
$(i,j)$ & HV & NDR \\ \hline\hline
\V$(7,3)$ & $\left(-3N+\frac{8}{3N}\right)\left(u-\frac{1}{2}d\right)$ &
$\left(-3N+\frac{20}{3N}\right)\left(u-\frac{1}{2}d\right)$ \\
\V$(7,4)$ & $\frac{1}{3}\left(u-\frac{1}{2}d\right)$ &
$-\frac{11}{3}\left(u-\frac{1}{2}d\right)$ \\
\V$(7,5)$ & $\left(3N-\frac{10}{3N}\right)\left(u-\frac{1}{2}d\right)$ &
$\left(3N+\frac{2}{3N}\right)\left(u-\frac{1}{2}d\right)$ \\
\V$(7,6)$ & $\frac{1}{3}\left(u-\frac{1}{2}d\right)$ &
$-\frac{11}{3}\left(u-\frac{1}{2}d\right)$ \\
\V$(7,7)$ & $\frac{44N^2}{3}-\frac{71}{6}+\frac{15}{2N^2}-\frac{8N}{3}f+
\frac{10}{3N}f$ & $\frac{137}{6}+\frac{15}{2N^2}-\frac{22}{3N}f$\\
\V$(7,8)$ & $-\frac{40N}{3}+\frac{3}{N}-\frac{2}{3}f$ & $-\frac{100N}{3}
+\frac{3}{N}+\frac{22}{3}f$ \\[0.5ex]
\hline
\V$(8,3)$ & $\left(-\frac{128}{27}-\frac{94}{27N^2}\right)
\left(u-\frac{1}{2}d\right)$ & $\left(-\frac{56}{27}-\frac{178}{27N^2}
\right)\left(u-\frac{1}{2}d\right)$ \\
\V$(8,4)$ & $\left(\frac{20N}{27}+\frac{202}{27N}\right)
\left(u-\frac{1}{2}d\right)$ & $\left(-\frac{16N}{27}+\frac{250}{27N}
\right)\left(u-\frac{1}{2}d\right)$ \\
\V$(8,5)$ & $\left(-\frac{38}{27}+\frac{86}{27N^2}\right)
\left(u-\frac{1}{2}d\right)$ & $\left(\frac{70}{27}+\frac{74}{27N^2}
\right)\left(u-\frac{1}{2}d\right)$ \\
\V$(8,6)$ & $\left(\frac{110N}{27}-\frac{158}{27N}\right)
\left(u-\frac{1}{2}d\right)$ & $\left(\frac{110N}{27}-\frac{254}{27N}
\right)\left(u-\frac{1}{2}d\right)$ \\
\V$(8,7)$ & $\frac{107N}{6}-\frac{18}{N}-\frac{4}{3}f$ & $-\frac{71N}{2}
-\frac{18}{N}+4f$ \\
\V$(8,8)$ & $-\frac{9N^2}{2}-\frac{17}{6}+\frac{15}{2N^2}-2Nf+\frac{10}{3N}f$ &
$\frac{-203N^2+479}{6}+\frac{15}{2N^2}+\frac{10N}{3}f-\frac{22}{3N}f$ \\[0.5ex]
\hline
\V$(9,3)$ & $\frac{56}{27}-\frac{86}{27N^2}+\left(3N-\frac{4}{N}\right)
\left(u-\frac{1}{2}d\right)$ & $\frac{32}{27}-\frac{86}{27N^2}
+\left(3N-\frac{8}{3N}\right)\left(u-\frac{1}{2}d\right)$ \\
\V$(9,4)$ & $-\frac{110N}{27}+\frac{140}{27N}+u-\frac{1}{2}d$ &
$-\frac{176N}{27}
+\frac{230}{27N}-\frac{1}{3}\left(u-\frac{1}{2}d\right)$ \\
\V$(9,5)$ & $\frac{128}{27}+\frac{58}{27N^2}+$ &
$\frac{122}{27}+\frac{94}{27N^2}+$ \\
\V & $\left(-3N+\frac{2}{N}\right)\left(u-\frac{1}{2}d\right)$ &
$\left(-3N+\frac{10}{3N}\right)\left(u-\frac{1}{2}d\right)$ \\
\V$(9,6)$ & $-\frac{38N}{27}-\frac{148}{27N}+u-\frac{1}{2}d$ & $-\frac{86N}{27}
-\frac{130}{27N}-\frac{1}{3}\left(u-\frac{1}{2}d\right)$ \\
\V$(9,9)$ & $\frac{44N^2}{3}-\frac{110}{3}-\frac{57}{2N^2}-\frac{8N}{3}f+
\frac{14}{3N}f$ & $-\frac{22}{3}-\frac{57}{2N^2}-\frac{2}{3N}f$ \\
\V$(9,10)$ & $\frac{23N}{2}+\frac{39}{N}-2f$ & $-\frac{19N}{6}+\frac{39}{N}
+\frac{2}{3}f$ \\[0.5ex]
\hline
\V$(10,3)$ & $-3N+\frac{4}{N}+$ & $-3N+\frac{2}{3N}+$ \\
\V & $\left(-\frac{56}{27}+\frac{86}{27N^2}\right)\left(u-\frac{1}{2}d\right)$
&
$\left(-\frac{20}{27}+\frac{74}{27N^2}\right)\left(u-\frac{1}{2}d\right)$ \\
\V$(10,4)$ & $-1+\left(\frac{110N}{27}-\frac{140}{27N}\right)
\left(u-\frac{1}{2}d\right)$ & $\frac{7}{3}+\left(\frac{110N}{27}
-\frac{164}{27N}\right)\left(u-\frac{1}{2}d\right)$ \\
\V$(10,5)$ & $\frac{3N^2-2}{N}+\left(-\frac{128}{27}-\frac{58}{27N^2}\right)
\left(u-\frac{1}{2}d\right)$ & $\frac{9N^2-16}{3N}+\left(-\frac{56}{27}
+\frac{2}{27N^2}\right)\left(u-\frac{1}{2}d\right)$ \\
\V$(10,6)$ & $-1+\left(\frac{38N}{27}+\frac{148}{27N}\right)
\left(u-\frac{1}{2}d\right)$ & $\frac{7}{3}+\left(\frac{74N}{27}-\frac{20}{27N}
\right)\left(u-\frac{1}{2}d\right)$ \\
\V$(10,9)$ & $\frac{23N}{2}+\frac{39}{N}-2f$ & $-\frac{19N}{6}+\frac{39}{N}
+\frac{2}{3}f$ \\
\V$(10,10)$ & $\frac{44N^2}{3}-\frac{110}{3}-\frac{57}{2N^2}-\frac{8N}{3}f
+\frac{14}{3N}f$ & $-\frac{22}{3}-\frac{57}{2N^2}-\frac{2}{3N}f$ \\[0.5ex]
\hline
\end{tabular}
\caption[]{(Continued) Elements of the two-loop anomalous dimension matrix
$\left(\hat\gamma_{\sss s}^{\sss (1)}\right)_{ij}$.}
\end{center}
\end{table}

\newpage
\clearpage

\begin{table}
\begin{center}
\begin{tabular}{|c|c|c|}\hline
$(i,j)$ & HV & NDR \\
\hline\hline
\V$(1,1)$ & $8N-\frac{22}{3N}$ & $8N-\frac{22}{3N}$ \\
\V$(1,2)$ & $-\frac{2}{3}$ & $-\frac{2}{3}$ \\
\V$(1,3)$ & $-\frac{88}{243}$ & $-\frac{88}{243}$ \\
\V$(1,4)$ & $\frac{88N}{243}$ & $\frac{88N}{243}$ \\
\V$(1,5)$ & $-\frac{88}{243}$ & $-\frac{88}{243}$ \\
\V$(1,6)$ & $\frac{88N}{243}$ & $\frac{88N}{243}$ \\
\V$(1,7)$ & $\frac{8N^2}{9}-\frac{16}{27}$ & $\frac{8N^2}{9}-\frac{64}{27}$ \\
\V$(1,8)$ & $-\frac{8N}{27}$ & $\frac{40N}{27}$ \\
\V$(1,9)$ & $\frac{8N^2}{9}$ & $\frac{8N^2}{9}-\frac{80}{27}$ \\
\V$(1,10)$ & $-\frac{8N}{9}$ & $\frac{56N}{27}$ \\[0.5ex]
\hline
\V$(2,1)$ & $\frac{25}{3}$ & $\frac{25}{3}$ \\
\V$(2,2)$ & $-N-\frac{22}{3N}$ & $-N-\frac{22}{3N}$ \\
\V$(2,3)$ & $-\frac{268}{243N}$ & $-\frac{556}{243N}$ \\
\V$(2,4)$ & $\frac{268}{243}$ & $\frac{556}{243}$ \\
\V$(2,5)$ & $-\frac{268}{243N}$ & $-\frac{556}{243N}$ \\
\V$(2,6)$ & $\frac{268}{243}$ & $\frac{556}{243}$ \\
\V$(2,7)$ & $\frac{520N}{243}+\frac{884}{243N}$ & $-\frac{200N}{243}+
\frac{1316}{243N}$ \\
\V$(2,8)$ & $-\frac{52}{9}$ & $-\frac{124}{27}$ \\
\V$(2,9)$ & $\frac{520N}{243}-\frac{1636}{243N}$ & $-\frac{200N}{243}-
\frac{1348}{243N}$ \\
\V$(2,10)$ & $\frac{124}{27}$ & $\frac{172}{27}$ \\[0.5ex]
\hline
\V$(3,3)$ & $\frac{1402}{243N}-\frac{88}{243}\left(u-\frac{1}{2}d\right)$ &
$\frac{1690}{243N}-\frac{136}{243}\left(u-\frac{1}{2}d\right)$ \\
\V$(3,4)$ & $-\frac{1402}{243}+\frac{88N}{243}\left(u-\frac{1}{2}d\right)$ &
$-\frac{1690}{243}+\frac{136N}{243}\left(u-\frac{1}{2}d\right)$ \\
\V$(3,5)$ & $-\frac{56}{243N}-\frac{88}{243}\left(u-\frac{1}{2}d\right)$ &
$\frac{232}{243N}-\frac{136}{243}\left(u-\frac{1}{2}d\right)$ \\
\V$(3,6)$ & $\frac{56}{243}+\frac{88N}{243}\left(u-\frac{1}{2}d\right)$ &
$-\frac{232}{243}+\frac{136N}{243}\left(u-\frac{1}{2}d\right)$ \\
\V$(3,7)$ & $-\frac{256N}{243}-\frac{1040}{243N}+$ & $\frac{464N}{243}-
\frac{1040}{243N}+$ \\
\V & $\left(\frac{8N^2}{9}-\frac{16}{27}\right)\left(u-\frac{1}{2}d\right)$ &
$\left(\frac{8N^2}{9}-\frac{112}{27}\right)\left(u-\frac{1}{2}d\right)$ \\
\V$(3,8)$ & $\frac{16}{3}-\frac{8N}{27}\left(u-\frac{1}{2}d\right)$ &
$\frac{64}{27}+\frac{88N}{27}\left(u-\frac{1}{2}d\right)$ \\
\V$(3,9)$ & $\frac{1688N}{243}-\frac{1976}{243N}+$ & $\frac{2408N}{243}
-\frac{1400}{243N}+$ \\
\V & $\frac{8N^2}{9}\left(u-\frac{1}{2}d\right)$ &
$\left(\frac{8N^2}{9}-\frac{32}{27}\right)\left(u-\frac{1}{2}d\right)$ \\
\V$(3,10)$ & $\frac{32}{27}-\frac{8N}{9}\left(u-\frac{1}{2}d\right)$ &
$-\frac{112}{27}+\frac{8N}{27}\left(u-\frac{1}{2}d\right)$ \\[0.5ex]
\hline
\end{tabular}
\caption[]{Elements of the two-loop anomalous dimension matrix
$\left(\hat\gamma_{\sss e}^{\sss (1)}\right)_{ij}$.
The elements which are not shown are equal to zero. (Continue)}
\end{center}
\end{table}

\newpage
\clearpage

\addtocounter{table}{-1}
\begin{table}
\begin{center}
\begin{tabular}{|c|c|c|}\hline
$(i,j)$ & HV & NDR \\
\hline \hline
\V$(4,3)$ &
$-\frac{641}{243}+\frac{8}{27N}f-\frac{340}{243N}u-\frac{100}{243N}d$ &
$-\frac{641}{243}-\frac{388}{243N}u+\frac{32}{243N}d$ \\
\V$(4,4)$ & $-\frac{817N}{243}+\frac{6}{N}-\frac{8}{27}f+\frac{340}{243}u+
\frac{100}{243}d$ & $-\frac{817N}{243}+\frac{6}{N}+\frac{388}{243}u-
\frac{32}{243}d$ \\
\V$(4,5)$ & $\frac{88}{243}+\frac{8}{27N}f-\frac{340}{243N}u-\frac{100}{243N}d$
&
$\frac{88}{243}-\frac{388}{243N}u+\frac{32}{243N}d$ \\
\V$(4,6)$ & $-\frac{88N}{243}-\frac{8}{27}f+\frac{340}{243}u+\frac{100}{243}d$
&
$-\frac{88N}{243}+\frac{388}{243}u-\frac{32}{243}d$ \\
\V$(4,7)$ & $-\frac{8N^2}{9}+\frac{16}{27}+\frac{88N}{243}f-\frac{52}{243N}f+$
&
$-\frac{8N^2}{9}+\frac{64}{27}+\frac{136N}{243}f+\frac{260}{243N}f+$ \\
\V & $\left(\frac{16N}{9}+\frac{104}{27N}\right)\left(u-\frac{1}{2}d\right)$ &
$\left(\frac{16N}{27}+\frac{40}{27N}\right)\left(u-\frac{1}{2}d\right)$ \\
\V$(4,8)$ &
$\frac{8N}{27}-\frac{4}{27}f-\frac{152}{27}\left(u-\frac{1}{2}d\right)$ &
$-\frac{40N}{27}-\frac{44}{27}f-\frac{56}{27}\left(u-\frac{1}{2}d\right)$ \\
\V$(4,9)$ & $\frac{34}{3}-\frac{8N^2}{9}+\frac{88N}{243}f-\frac{124}{243N}f+$ &
$-\frac{8N^2}{9}+\frac{386}{27}+\frac{136N}{243}f-\frac{100}{243N}f+$ \\
\V & $\left(\frac{16N}{9}-\frac{56}{9N}\right)\left(u-\frac{1}{2}d\right)$ &
$\left(\frac{16N}{27}-\frac{56}{9N}\right)\left(u-\frac{1}{2}d\right)$ \\
\V$(4,10)$ & $\frac{26N}{9}-\frac{40}{3N}+\frac{4}{27}f+\frac{40}{9}
\left(u-\frac{1}{2}d\right)$ & $-\frac{2N}{27}-\frac{40}{3N}-\frac{4}{27}f
+\frac{152}{27}\left(u-\frac{1}{2}d\right)$ \\[0.5ex]
\hline
\V$(5,3)$ & $-\frac{88}{243}\left(u-\frac{1}{2}d\right)$ &
$-\frac{136}{243}\left(u-\frac{1}{2}d\right)$ \\
\V$(5,4)$ & $\frac{88N}{243}\left(u-\frac{1}{2}d\right)$ &
$\frac{136N}{243}\left(u-\frac{1}{2}d\right)$ \\
\V$(5,5)$ & $-\frac{6}{N}-\frac{88}{243}\left(u-\frac{1}{2}d\right)$ &
$-\frac{6}{N}-\frac{136}{243}\left(u-\frac{1}{2}d\right)$ \\
\V$(5,6)$ & $6+\frac{88N}{243}\left(u-\frac{1}{2}d\right)$ &
$6+\frac{136N}{243}\left(u-\frac{1}{2}d\right)$ \\
\V$(5,7)$ & $-8N-\frac{16}{3N}+$ & $-8N-\frac{16}{3N}+$ \\
\V & $\left(\frac{8N^2}{9}-\frac{16}{27}\right)\left(u-\frac{1}{2}d\right)$ &
$\left(\frac{8N^2}{9}-\frac{112}{27}\right)\left(u-\frac{1}{2}d\right)$\\
\V$(5,8)$ & $\frac{40}{3}-\frac{8N}{27}\left(u-\frac{1}{2}d\right)$ &
$\frac{40}{3}+\frac{88N}{27}\left(u-\frac{1}{2}d\right)$ \\
\V$(5,9)$ & $\frac{8N^2}{9}\left(u-\frac{1}{2}d\right)$ &
$\left(\frac{8N^2}{9}-\frac{32}{27}\right)\left(u-\frac{1}{2}d\right)$ \\
\V$(5,10)$ & $-\frac{8N}{9}\left(u-\frac{1}{2}d\right)$ &
$\frac{8N}{27}\left(u-\frac{1}{2}d\right)$ \\[0.5ex]
\hline
\V$(6,3)$ & $\frac{8}{27N}f-\frac{412}{243N}u-\frac{64}{243N}d$ &
$-\frac{748}{243N}u+\frac{212}{243N}d$ \\
\V$(6,4)$ & $-\frac{8}{27}f+\frac{412}{243}u+\frac{64}{243}d$ &
$\frac{748}{243}u-\frac{212}{243}d$ \\
\V$(6,5)$ & $3+\frac{8}{27N}f-\frac{412}{243N}u-\frac{64}{243N}d$ &
$3-\frac{748}{243N}u+\frac{212}{243N}d$ \\
\V$(6,6)$ & $3N-\frac{6}{N}-\frac{8}{27}f+\frac{412}{243}u+\frac{64}{243}d$ &
$3N-\frac{6}{N}+\frac{748}{243}u-\frac{212}{243}d$ \\
\V$(6,7)$ & $-2+\frac{88N}{243}f-\frac{52}{243N}f+$ &
$-2+\frac{136N}{243}f+\frac{260}{243N}f-$ \\
\V & $\left(\frac{32N}{27}-\frac{56}{9N}\right)\left(u-\frac{1}{2}d\right)$ &
$\left(\frac{64N}{27}+\frac{56}{9N}\right)\left(u-\frac{1}{2}d\right)$ \\
\V$(6,8)$ & $\frac{22N}{3}-\frac{16}{3N}-\frac{4}{27}f+
\frac{136}{27}\left(u-\frac{1}{2}d\right)$ &
$\frac{22N}{3}-\frac{16}{3N}-\frac{44}{27}f+\frac{232}{27}
\left(u-\frac{1}{2}d\right)$ \\
\V$(6,9)$ & $\frac{88N}{243}f-\frac{124}{243N}f+$ &
$\frac{136N}{243}f-\frac{100}{243N}f+$ \\
\V & $\left(\frac{32N}{27}+\frac{136}{27N}\right)\left(u-\frac{1}{2}d\right)$ &
$\left(-\frac{64N}{27}+\frac{200}{27N}\right)\left(u-\frac{1}{2}d\right)$ \\
\V$(6,10)$ & $\frac{4}{27}f-\frac{56}{9}\left(u-\frac{1}{2}d\right)$ &
$-\frac{4}{27}f-\frac{136}{27}\left(u-\frac{1}{2}d\right)$ \\[0.5ex]
\hline
\end{tabular}
\caption[]{(Continued) Elements of the two-loop anomalous dimension matrix
$\left(\hat\gamma_{\sss e}^{\sss (1)}\right)_{ij}$ (Continue)}
\end{center}
\end{table}

\newpage
\clearpage

\addtocounter{table}{-1}
\begin{table}
\begin{center}
\begin{tabular}{|c|c|c|}\hline
$(i,j)$ & HV & NDR \\ \hline\hline
\V$(7,3)$ & $-\frac{88}{243}\left(u+\frac{1}{4}d\right)$ &
$-\frac{136}{243}\left(u+\frac{1}{4}d\right)$ \\
\V$(7,4)$ & $\frac{88N}{243}\left(u+\frac{1}{4}d\right)$ &
$\frac{136N}{243}\left(u+\frac{1}{4}d\right)$ \\
\V$(7,5)$ & $-4N-\frac{8}{3N}-\frac{88}{243}\left(u+\frac{1}{4}d\right)$ &
$-4N-\frac{8}{3N}-\frac{136}{243}\left(u+\frac{1}{4}d\right)$ \\
\V$(7,6)$ & $\frac{20}{3}+\frac{88N}{243}\left(u+\frac{1}{4}d\right)$ &
$\frac{20}{3}+\frac{136N}{243}\left(u+\frac{1}{4}d\right)$ \\
\V$(7,7)$ & $-4N-\frac{26}{3N}+$ & $-4N-\frac{26}{3N}+$ \\
\V & $\left(\frac{8N^2}{9}-\frac{16}{27}\right)\left(u+\frac{1}{4}d\right)$ &
$\left(\frac{8N^2}{9}-\frac{112}{27}\right)\left(u+\frac{1}{4}d\right)$\\
\V$(7,8)$ & $\frac{38}{3}-\frac{8N}{27}\left(u+\frac{1}{4}d\right)$ &
$\frac{38}{3}+\frac{88N}{27}\left(u+\frac{1}{4}d\right)$ \\
\V$(7,9)$ & $\frac{8N^2}{9}\left(u+\frac{1}{4}d\right)$ &
$\left(\frac{8N^2}{9}-\frac{32}{27}\right)\left(u+\frac{1}{4}d\right)$ \\
\V$(7,10)$ & $-\frac{8N}{9}\left(u+\frac{1}{4}d\right)$ &
$\frac{8N}{27}\left(u+\frac{1}{4}d\right)$ \\[0.5ex]
\hline
\V$(8,3)$ & $-\frac{340}{243N}u-\frac{4}{243N}d$ &
$-\frac{748}{243N}u-\frac{106}{243N}d$ \\
\V$(8,4)$ & $\frac{340}{243}u+\frac{4}{243}d$ &
$\frac{748}{243}u+\frac{106}{243}d$ \\
\V$(8,5)$ & $-1-\frac{340}{243N}u-\frac{4}{243N}d$ &
$-1-\frac{748}{243N}u-\frac{106}{243N}d$ \\
\V$(8,6)$ & $\frac{11N}{3}-\frac{8}{3N}+\frac{340}{243}u+\frac{4}{243}d$ &
$\frac{11N}{3}-\frac{8}{3N}+\frac{748}{243}u+\frac{106}{243}d$ \\
\V$(8,7)$ & $2+\frac{376N}{243}u-\frac{1564}{243N}u+$ &
$2-\frac{440N}{243}u-\frac{1252}{243N}u-$ \\
\V &$\frac{28N}{243}d-\frac{352}{243N}d$ &
$\frac{212N}{243}d-\frac{508}{243N}d$ \\
\V$(8,8)$ & $\frac{20N}{3}-\frac{26}{3N}+\frac{44}{9}u+\frac{4}{3}d$ &
$\frac{20N}{3}-\frac{26}{3N}+\frac{188}{27}u+\frac{80}{27}d$ \\
\V$(8,9)$ & $\frac{376N}{243}u+\frac{1100}{243N}u+\frac{28N}{243}d+
\frac{368}{243N}d$ & $-\frac{440N}{243}u+\frac{1700}{243N}u-\frac{212N}{243}d
+\frac{500}{243N}d$ \\
\V$(8,10)$ & $-\frac{164}{27}u-\frac{44}{27}d$ & $-\frac{140}{27}u
-\frac{32}{27}d$ \\[0.5ex]
\hline
\V$(9,3)$ & $4N-\frac{1592}{243N}-\frac{88}{243}\left(u+\frac{1}{4}d\right)$ &
$4N-\frac{1736}{243N}-\frac{136}{243}\left(u+\frac{1}{4}d\right)$ \\
\V$(9,4)$ & $\frac{620}{243}+\frac{88N}{243}\left(u+\frac{1}{4}d\right)$ &
$\frac{764}{243}+\frac{136N}{243}\left(u+\frac{1}{4}d\right)$ \\
\V$(9,5)$ & $\frac{28}{243N}-\frac{88}{243}\left(u+\frac{1}{4}d\right)$ &
$-\frac{116}{243N}-\frac{136}{243}\left(u+\frac{1}{4}d\right)$ \\
\V$(9,6)$ & $-\frac{28}{243}+\frac{88N}{243}\left(u+\frac{1}{4}d\right)$ &
$\frac{116}{243}+\frac{136N}{243}\left(u+\frac{1}{4}d\right)$ \\
\V$(9,7)$ & $\frac{128N}{243}+\frac{520}{243N}+$ &
$-\frac{232N}{243}+\frac{520}{243N}$\\
\V & $\left(\frac{8N^2}{9}-\frac{16}{27}\right)\left(u+\frac{1}{4}d\right)$ &
$+\left(\frac{8N^2}{9}-\frac{112}{27}\right)\left(u+\frac{1}{4}d\right)$ \\
\V$(9,8)$ & $-\frac{8}{3}-\frac{8N}{27}\left(u+\frac{1}{4}d\right)$ &
$-\frac{32}{27}+\frac{88N}{27}\left(u+\frac{1}{4}d\right)$ \\
\V$(9,9)$ & $\frac{1100N}{243}-\frac{794}{243N}+\frac{8N^2}{9}
\left(u+\frac{1}{4}d\right)$ & $\frac{740N}{243}-\frac{1082}{243N}+
\left(\frac{8N^2}{9}-\frac{32}{27}\right)\left(u+\frac{1}{4}d\right)$ \\
\V$(9,10)$ & $-\frac{34}{27}-\frac{8N}{9}\left(u+\frac{1}{4}d\right)$ &
$\frac{38}{27}+\frac{8N}{27}\left(u+\frac{1}{4}d\right)$ \\[0.5ex]
\hline
\end{tabular}
\caption[]{(Continued) Elements of the two-loop anomalous dimension matrix
$\left(\hat\gamma_{\sss e}^{\sss (1)}\right)_{ij}$ (Continue)}
\end{center}
\end{table}

\newpage
\clearpage

\addtocounter{table}{-1}
\begin{table}[ht]
\begin{center}
\begin{tabular}{|c|c|c|}\hline
$(i,j)$ & HV & NDR \\ \hline\hline
\V$(10,3)$ & $\frac{1333}{243}-\frac{268}{243N}u+\frac{14}{243N}d$ &
$\frac{1333}{243}-\frac{388}{243N}u-\frac{16}{243N}d$ \\
\V$(10,4)$ & $\frac{287N}{243}-\frac{20}{3N}+\frac{268}{243}u-\frac{14}{243}d$
&
$\frac{287N}{243}-\frac{20}{3N}+\frac{388}{243}u+\frac{16}{243}d$ \\
\V$(10,5)$ & $-\frac{44}{243}-\frac{268}{243N}u+\frac{14}{243N}d$ &
$-\frac{44}{243}-\frac{388}{243N}u-\frac{16}{243N}d$ \\
\V$(10,6)$ & $\frac{44N}{243}+\frac{268}{243}u-\frac{14}{243}d$ &
$\frac{44N}{243}+\frac{388}{243}u+\frac{16}{243}d$ \\
\V$(10,7)$ & $\frac{4N^2}{9}-\frac{8}{27}+\frac{520N}{243}u+$ &
$\frac{4N^2}{9}-\frac{32}{27}+\frac{280N}{243}u+$ \\
\V & $\frac{884}{243N}u+\frac{64N}{243}d+\frac{260}{243N}d$ &
$\frac{620}{243N}u-\frac{32N}{243}d-\frac{40}{243N}d$ \\
\V$(10,8)$ & $-\frac{4N}{27}-\frac{52}{9}u-\frac{4}{3}d$ &
$\frac{20N}{27}-\frac{100}{27}u+\frac{8}{27}d$ \\
\V$(10,9)$ & $\frac{4N^2}{9}+\frac{8}{3}+\frac{520N}{243}u-$
& $\frac{4N^2}{9}+\frac{32}{27}+\frac{280N}{243}u-$ \\
\V & $\frac{1636}{243N}u+\frac{64N}{243}d-\frac{316}{243N}d$ &
$\frac{1612}{243N}u-\frac{32N}{243}d-\frac{328}{243N}d$ \\
\V$(10,10)$ & $-\frac{22N}{9}-\frac{2}{3N}+\frac{124}{27}u+\frac{28}{27}d$ &
$-\frac{26N}{27}-\frac{2}{3N}+\frac{148}{27}u+\frac{40}{27}d$ \\[0.5ex]
\hline
\end{tabular}
\caption[]{(Continued) Elements of the two-loop anomalous dimension matrix
$\left(\hat\gamma_{\sss e}^{\sss (1)}\right)_{ij}$}
\end{center}
\end{table}

\newpage
\clearpage

\section*{Acknowledgments}
\par We thank P. Nason for his participation to the earlier stages
of this work. We also thank G. Altarelli,
A. Buras, M. Jamin, M. Lusignoli, L. Maiani, M. Veltman and P. Weisz for many
interesting discussions on the subject of this paper. We acknowledge
the partial support of the MURST, Italy, and INFN.

\section*{Figure Captions}
\begin{itemize}
\item fig.1: Tree-level current-current diagram.
\item fig.2a-c: One-loop $O(\alphas)$ current-current diagrams in
the full theory.
\item fig.3a-c: One-loop $O(\alphas)$ current-current diagrams in
the effective  theory. In the figure the relevant momenta of
the external legs are given.
\item fig.4: One-loop $O(\alphas)$  quark self-energy diagram.
\item fig.5a-d: One loop $O(\alphae)$ current-current diagrams in
the full theory, including the non-abelian  diagrams.
The $Z^0$-$W$ diagrams are also shown.
\item fig.6a-b: The quark and $W$ self-energy diagrams at $O(\alphae)$.
\item fig.7: Box diagrams contributing to $B(\xt)$.
\item fig.8a-b: Penguin diagrams in the full theory.
 In the electro-weak case also the non-abelian diagram is shown.
\item fig.9: Penguin diagrams in the effective theory.
 In the figure the relevant momenta of
the external legs are given.
\item fig.10: Current-current diagrams at two loops.
\item fig.11: Penguin diagrams at two loops.
\item fig.12:  Schematic representation of the counter-terms: current-current
 (a), one-gluon penguin (b) and
two-gluon counterterms (c).
\item fig.13: Diagramatic representation of the subtraction procedure
for a current-current diagram.
The ``complete" diagram is obtained by summing the diagram with a bare operator
inserted, the counter-terms including those corresponding to
effervescent operators and the term defined as $E_{LL}^{(17)}$
in eq.(\ref{fri}).
\item fig.14:Diagramatic representation of the subtraction procedure
for a penguin diagram.
The ``complete" diagram is obtained by summing the diagram with a bare operator
inserted, the counter-terms including those corresponding to
effervescent operators and the term defined as $E_{LL}^{(3)}$ in
eq.(\ref{fro}).
Also the two-gluon counter-term is shown.
\item fig.15: One loop diagram for the
$s \rightarrow d + g + g \,\,\, \,\, ( g + \gamma)$ operator. From
this diagram one can compute $p_{2gct}$ and $\Delta r_{2gct}$,
cf. eq.(\ref{pippo}).
\item fig.16: This figure shows that  the sum of diagrams $P_2$ and $P_3$ is
the same with or without the two-gluon counter-terms. This corresponds
to the cancellation of counter-terms which vanish by the equations of motion
in the abelian case.
\item fig.17: The contribution of the longitudinal term
 $\sim q^{\mu}q\!\!\!/$ of
the penguin counter-term cancels when we sum the diagrams shown in this
figure. This corresponds to the cancellation of counter-terms which vanish by
the equations of motion in the abelian case.
\item fig.18: Quark self-energy diagrams at two loops.
\end{itemize}
\newpage
\clearpage

\setcounter{table}{4}

\section*{Appendix}
\label{app:A}

\begin{table}[h]
\begin{center}
\begin{tabular}{|c|c|c|}\hline\hline
Diagram & \tsq$\frac{1}{\epuno}$ & $O(1)$ \\ \hline
$S_0$ & -1 & -1/2 \\
$S_1$ & 1/8 & $-$  \\
$S_2$ &  $f$/2 & $-$ \\
$S_3$ &  5/8 & $-$ \\
$S_4$ &  1/4 & $-$ \\
$S_5$ &  11/4 & $-$ \\ \hline\hline
\end{tabular}
\caption[]{Single pole and finite part for the
one-loop and two-loop self-energy diagrams in fig. 4, 6 and 18. For
the two-loop case only the pole part is given.}
\label{1and2loopSE}
\end{center}
\end{table}

\newpage
\clearpage

\small

\begin{table}
\begin{center}
\begin{tabular}{|c|c|ccc|ccc|cc|ccc|}
\hline\hline
\multicolumn{1}{|c} { } & \multicolumn{1}{c}{ } &
\multicolumn{3}{|c}{$D$} &
\multicolumn{3}{|c}{$C$} &
\multicolumn{2}{|c}{$E$} &
\multicolumn{3}{|c|}{\tsq$\overline{D}$} \\
\hline
\multicolumn{1}{|c|}{$T_N$}& $M$ &
$\frac{1}{\epdue}$ & $\tsq{\frac{1}{\epuno}}^{\sss HV}$ &
${\frac{1}{\epuno}}^{\sss NDR}$ & $\frac{1}{\epdue}$ &
${\frac{1}{\epuno}}^{\sss HV}$ & ${\frac{1}{\epuno}}^{\sss NDR}$ &
${\frac{1}{\epuno}}^{\sss HV}$ & ${\frac{1}{\epuno}}^{\sss NDR}$ &
$\frac{1}{\epdue}$ & ${\frac{1}{\epuno}}^{\sss HV}$ &
${\frac{1}{\epuno}}^{\sss NDR}$ \\[0.5ex]
\hline
\tsp$V_4$ & 2 & $\frac{1}{2}$ & $\frac{15}{4}$ & $\frac{7}{4}$ &
1 & $\frac{5}{2}$ & $\frac{1}{2}$ & $-$ & $-$ & -$\frac{1}{2}$ &
$\frac{5}{4}$ & $\frac{5}{4}$ \\
\tsp$V_5$ & 2 & 8 & 37 & 41 & 16 & 22 & 30 & -7 & -3 & -8 & 8 & 8 \\
\tsp$V_6$ & 2 & $\frac{1}{2}$ & $\frac{15}{4}$ & $\frac{7}{4}$ &
1 & $\frac{5}{2}$ & $\frac{1}{2}$ & $-$ & $-$ & -$\frac{1}{2}$ &
$\frac{5}{4}$ & $\frac{5}{4}$ \\
\tsp$V_7$ & 2 & $-$ & -2 & -2 &
$-$ & $-$ & $-$ & $-$ & $-$ & $-$ & -2 & -2 \\
\tsp$V_8$ & 2 & $-$ & -2 & -2 &
$-$ & $-$ & $-$ & $-$ & $-$ & $-$ & -2 & -2 \\
\tsp$V_9$ & 2 & $-$ & -2 & -2 &
$-$ & $-$ & $-$ & $-$ & $-$ & $-$ & -2 & -2 \\
\tsp$V_{10}$ & 4 & $\frac{1}{2}$ & $\frac{9}{4}$ & $\frac{5}{4}$ &
1 & $\frac{5}{2}$ & $\frac{1}{2}$ & $-$ & $-$ & -$\frac{1}{2}$ &
-$\frac{1}{4}$ & $\frac{3}{4}$ \\
\tsp$V_{11}$ & 4 & -2 & -$\frac{11}{2}$ & -$\frac{11}{2}$ &
-4 & -9 & -9 & $-$ & $-$ & 2 & $\frac{7}{2}$ & $\frac{7}{2}$ \\
\tsp$V_{12}$ & 4 & $\frac{1}{2}$ & $\frac{9}{4}$ & $\frac{5}{4}$ &
1 & $\frac{5}{2}$ & $\frac{1}{2}$ & $-$ & $-$ & -$\frac{1}{2}$ &
-$\frac{1}{4}$ & $\frac{3}{4}$ \\
\tsp$V_{13}$ & 4 & -$\frac{1}{2}$ & -$\frac{9}{4}$ & -$\frac{5}{4}$ &
-1 & -$\frac{5}{2}$ & -$\frac{1}{2}$ & $-$ & $-$ & $\frac{1}{2}$ &
$\frac{1}{4}$ & -$\frac{3}{4}$ \\
\tsp$V_{14}$ & 4 & 2 & $\frac{17}{2}$ & $\frac{17}{2}$ &
4 & 9 & 9 & $-$ & $-$ & -2 & -$\frac{1}{2}$ & -$\frac{1}{2}$ \\
\tsp$V_{15}$ & 4 & -$\frac{1}{2}$ & -$\frac{9}{4}$ & -$\frac{5}{4}$ &
-1 & -$\frac{5}{2}$ & -$\frac{1}{2}$ & $-$ & $-$ & $\frac{1}{2}$ &
$\frac{1}{4}$ & -$\frac{3}{4}$ \\
\tsp$V_{16}$ & 4 & -2 & -$\frac{17}{2}$ & -$\frac{17}{2}$ &
-4 & -8 & -8 & 1 & -3 & 2 & $\frac{1}{2}$ & -$\frac{7}{2}$ \\
\tsp$V_{17}$ & 4 & -2 & -$\frac{19}{2}$ & -$\frac{11}{2}$ &
-4 & -9 & -9 & $-$ & $-$ & 2 & -$\frac{1}{2}$ & $\frac{7}{2}$ \\
\tsp$V_{18}$ & 4 & $\frac{1}{2}$ & $\frac{17}{4}$ & -$\frac{7}{4}$ &
1 & $\frac{9}{2}$ & -$\frac{11}{2}$ & 1 & -3 & -$\frac{1}{2}$ &
$\frac{3}{4}$ & $\frac{3}{4}$ \\
\tsp$V_{19}$ & 4 & $\frac{1}{2}$ & $\frac{13}{4}$ & $\frac{5}{4}$ &
1 & $\frac{5}{2}$ & $\frac{1}{2}$ & $-$ & $-$ & -$\frac{1}{2}$ &
$\frac{3}{4}$ & $\frac{3}{4}$ \\
\tsp$V_{20}$ & 4 & -2 & -$\frac{17}{2}$ & -$\frac{17}{2}$ &
-4 & -7 & -15 & 1 & -3 & 2 & -$\frac{1}{2}$ & $\frac{7}{2}$ \\
\tsp$V_{21}$ & 4 & -2 & -$\frac{19}{2}$ & -$\frac{11}{2}$ &
-4 & -10 & -2 & $-$ & $-$ & 2 & $\frac{1}{2}$ & -$\frac{7}{2}$ \\
\tsp$V_{22}$ & 1 & 1 & 5 & 1 & 2 & 5 & 1 & $-$ & $-$ & -1 & $-$ & $-$ \\
\tsp$V_{23}$ & 1 & 16 & 74 & 66 & 32 & 76 & 60 & 2 & -6 & -16 & $-$ & $-$ \\
\tsp$V_{24}$ & 1 & 1 & 5 & 1 & 2 & 5 & 1 & $-$ & $-$ & -1 & $-$ & $-$ \\
\tsp$V_{25}$ & 4 & -$\frac{3}{2}$ & -$\frac{29}{4}$ & -$\frac{17}{4}$ &
-3 & -$\frac{15}{2}$ & -$\frac{3}{2}$ & $-$ & $-$ & $\frac{3}{2}$ &
$\frac{1}{4}$ & -$\frac{11}{4}$ \\
\tsp$V_{26}$ & 4 & 6 & $\frac{55}{2}$ & $\frac{55}{2}$ &
12 & 27 & 27 & $-$ & $-$ & -6 & $\frac{1}{2}$ & $\frac{1}{2}$ \\
\tsp$V_{27}$ & 4 & -$\frac{3}{2}$ & -$\frac{29}{4}$ & -$\frac{17}{4}$ &
-3 & -$\frac{15}{2}$ & -$\frac{3}{2}$ & $-$ & $-$ & $\frac{3}{2}$ &
$\frac{1}{4}$ & -$\frac{11}{4}$ \\
\tsp$V_{28}$ & 4 & $-$ & $-$ & $-$ & $-$ & $-$ & $-$ & $-$ & $-$ & $-$ &
$-$ & $-$ \\
\tsp$V_{29}^{N}$ & 2 & $-$ & $\frac{35}{12}$ & $\frac{5}{4}$ &
$-$ & $\frac{10}{3}$ & $-$ & $-$ & $-$ & $-$ &
-$\frac{5}{12}$ & $\frac{5}{4}$ \\
\tsp$V_{29}^{f}$ & 2 & $-$ & -$\frac{7}{6}$ & -$\frac{1}{2}$ & $-$ &
-$\frac{4}{3}$ & $-$ & $-$ & $-$ & $-$ & $\frac{1}{6}$ & -$\frac{1}{2}$ \\
\tsp$V_{30}^{N}$ & 2 & -$\frac{5}{2}$ & -$\frac{27}{2}$ & -$\frac{27}{2}$ &
-5 & -$\frac{35}{3}$ & -$\frac{35}{3}$ & $-$ & $-$ & $\frac{5}{2}$ &
-$\frac{11}{6}$ & -$\frac{11}{6}$ \\
\tsp$V_{30}^{f}$ & 2 & 1 & 5 & 5 & 2 & $\frac{14}{3}$ &
$\frac{14}{3}$ & $-$ & $-$ & -1 & $\frac{1}{3}$ & $\frac{1}{3}$ \\
\tsp$V_{31}^{N}$ & 2 & $-$ & $\frac{35}{12}$ & $\frac{5}{4}$ &
$-$ & $\frac{10}{3}$ & $-$ & $-$ & $-$ & $-$ &
-$\frac{5}{12}$ & $\frac{5}{4}$ \\
\tsp$V_{31}^{f}$ & 2 & $-$ & -$\frac{7}{6}$ & -$\frac{1}{2}$ & $-$ &
-$\frac{4}{3}$ & $-$ & $-$ & $-$ & $-$ & $\frac{1}{6}$ &
-$\frac{1}{2}$ \\[0.5ex]
\hline\hline
\end{tabular}
\caption[]{Two-loop pole contributions for the $\LL$ four-quark type
diagrams in fig. 10. For $V_{29}$, $V_{30}$ and $V_{31}$,
the results proportional to the number of colour $N$ ($V^{N}$) and
flavour $f$ ($V^{f}$) are separately reported.}
\label{vvll}
\end{center}
\end{table}

\newpage
\clearpage

\begin{table}
\begin{center}
\begin{tabular}{|c|c|ccc|ccc|cc|ccc|}
\hline\hline
\multicolumn{1}{|c} { } & \multicolumn{1}{c}{ } &
\multicolumn{3}{|c}{$D$} &
\multicolumn{3}{|c}{$C$} &
\multicolumn{2}{|c}{$E$} &
\multicolumn{3}{|c|}{\tsq$\overline{D}$} \\
\hline
\multicolumn{1}{|c|}{$T_N$}& $M$ &
$\frac{1}{\epdue}$ & $\tsq{\frac{1}{\epuno}}^{\sss HV}$ &
${\frac{1}{\epuno}}^{\sss NDR}$ & $\frac{1}{\epdue}$ &
${\frac{1}{\epuno}}^{\sss HV}$ & ${\frac{1}{\epuno}}^{\sss NDR}$ &
${\frac{1}{\epuno}}^{\sss HV}$ & ${\frac{1}{\epuno}}^{\sss NDR}$ &
$\frac{1}{\epdue}$ & ${\frac{1}{\epuno}}^{\sss HV}$ &
${\frac{1}{\epuno}}^{\sss NDR}$ \\[0.5ex]
\hline
\tsp$V_4$ & 2 & $\frac{1}{2}$ & $\frac{15}{4}$ & $\frac{7}{4}$ &
1 & $\frac{5}{2}$ & $\frac{1}{2}$ & $-$ & $-$ & -$\frac{1}{2}$ &
$\frac{5}{4}$ & $\frac{5}{4}$ \\
\tsp$V_5$ & 2 & $\frac{1}{2}$ & $\frac{7}{4}$ & $\frac{31}{4}$ & 1 &
-$\frac{1}{2}$ & $\frac{19}{2}$ & -1 & 3 & -$\frac{1}{2}$ & $\frac{5}{4}$ &
$\frac{5}{4}$ \\
\tsp$V_6$ & 2 & 8 & 48 & 32 &16 & 40 & 24 & $-$ & $-$ & -8 & 8 & 8 \\
\tsp$V_7$ & 2 & $-$ & -2 & -2 &
$-$ & $-$ & $-$ & $-$ & $-$ & $-$ & -2 & -2 \\
\tsp$V_8$ & 2 & $-$ & -2 & -2 &
$-$ & $-$ & $-$ & $-$ & $-$ & $-$ & -2 & -2 \\
\tsp$V_9$ & 2 & $-$ & -2 & -2 &
$-$ & $-$ & $-$ & $-$ & $-$ & $-$ & -2 & -2 \\
\tsp$V_{10}$ & 4 & $\frac{1}{2}$ & $\frac{9}{4}$ & $\frac{5}{4}$ &
1 & $\frac{5}{2}$ & $\frac{1}{2}$ & $-$ & $-$ & -$\frac{1}{2}$ &
-$\frac{1}{4}$ & $\frac{3}{4}$ \\
\tsp$V_{11}$ & 4 & -$\frac{1}{2}$ & -$\frac{7}{4}$ & -$\frac{11}{4}$ &
-1 & -$\frac{3}{2}$ & -$\frac{7}{2}$ & $-$ & $-$ & $\frac{1}{2}$ &
-$\frac{1}{4}$ & $\frac{3}{4}$ \\
\tsp$V_{12}$ & 4 & 2 & 6 & 4 & 4 & 10 & 6 & $-$ & $-$ & -2 & -4 & -2 \\
\tsp$V_{13}$ & 4 & -$\frac{1}{2}$ & -$\frac{9}{4}$ & -$\frac{5}{4}$ &
-1 & -$\frac{5}{2}$ & -$\frac{1}{2}$ & $-$ & $-$ & $\frac{1}{2}$ &
$\frac{1}{4}$ & -$\frac{3}{4}$ \\
\tsp$V_{14}$ & 4 & $\frac{1}{2}$ & $\frac{7}{4}$ & $\frac{11}{4}$ &
1 & $\frac{3}{2}$ & $\frac{7}{2}$ & $-$ & $-$ & -$\frac{1}{2}$ &
$\frac{1}{4}$ & -$\frac{3}{4}$ \\
\tsp$V_{15}$ & 4 & -2 & -9 & -7 & -4 & -10 & -6 & $-$ & $-$ & 2 & 1 & -1 \\
\tsp$V_{16}$ & 4 & -$\frac{1}{2}$ & -$\frac{7}{4}$ & $\frac{1}{4}$ &
-1 & -$\frac{1}{2}$ & $\frac{11}{2}$ & 1 & 3 & $\frac{1}{2}$ &
-$\frac{1}{4}$ & -$\frac{9}{4}$ \\
\tsp$V_{17}$ & 4 & -$\frac{1}{2}$ & -$\frac{11}{4}$ & -$\frac{11}{4}$ &
-1 & -$\frac{3}{2}$ & -$\frac{7}{2}$ & $-$ & $-$ & $\frac{1}{2}$ &
-$\frac{5}{4}$ & $\frac{3}{4}$ \\
\tsp$V_{18}$ & 4 & 2 & 11 & 7 & 4 & 12 & 8 & 1 & 3 & -2 & $-$ & 2 \\
\tsp$V_{19}$ & 4 & 2 & 10 & 4 & 4 & 10 & 6 & $-$ & $-$ & -2 & $-$ & -2 \\
\tsp$V_{20}$ & 4 & -2 & -7 & -7 & -4 & -4 & -8 & 1 & 3 & 2 & -2 & 4 \\
\tsp$V_{21}$ & 4 & -2 & -8 & -10 &-4 & -10 & -6 & $-$ & $-$ & 2 & 2 & -4 \\
\tsp$V_{22}$ & 1 & 1 & 5 & 1 & 2 & 5 & 1 & $-$ & $-$ & -1 & $-$ & $-$ \\
\tsp$V_{23}$ & 1 & 1 & 5 & 13 & 2 & 7 & 19 & 2 & 6 & -1 & $-$ & $-$ \\
\tsp$V_{24}$ & 1 & 16 & 80 & 48 & 32 & 80 & 48 & $-$ & $-$ & -16 & $-$ & $-$ \\
\tsp$V_{25}$ & 4 & -$\frac{3}{2}$ & -$\frac{29}{4}$ & -$\frac{17}{4}$ &
-3 & -$\frac{15}{2}$ & -$\frac{3}{2}$ & $-$ & $-$ & $\frac{3}{2}$ &
$\frac{1}{4}$ & -$\frac{11}{4}$ \\
\tsp$V_{26}$ & 4 & $\frac{3}{2}$ & $\frac{23}{4}$ & $\frac{35}{4}$ &
3 & $\frac{9}{2}$ & $\frac{21}{2}$ & $-$ & $-$ & -$\frac{3}{2}$ &
$\frac{5}{4}$ & -$\frac{7}{4}$ \\
\tsp$V_{27}$ & 4 & -6 & -29 & -23 & -12 & -30 & -18 & $-$ & $-$ & 6 & 1 & -5 \\
\tsp$V_{28}$ & 4 & $-$ & $-$ & $-$ & $-$ & $-$ & $-$ & $-$ & $-$ & $-$ &
$-$ & $-$ \\
\tsp$V_{29}^{N}$ & 2 & $-$ & $\frac{35}{12}$ & $\frac{5}{4}$ &
$-$ & $\frac{10}{3}$ & $-$ & $-$ & $-$ & $-$ &
-$\frac{5}{12}$ & $\frac{5}{4}$ \\
\tsp$V_{29}^{f}$ & 2 & $-$ & -$\frac{7}{6}$ & -$\frac{1}{2}$ & $-$ &
-$\frac{4}{3}$ & $-$ & $-$ & $-$ & $-$ & $\frac{1}{6}$ & -$\frac{1}{2}$ \\
\tsp$V_{30}^{N}$ & 2 & $-$ & -$\frac{25}{12}$ & -$\frac{15}{4}$ &
$-$ & -$\frac{5}{3}$ & -5 & $-$ & $-$ & $-$ &
-$\frac{5}{12}$ & $\frac{5}{4}$ \\
\tsp$V_{30}^{f}$ & 2 & $-$ & $\frac{5}{6}$ & $\frac{3}{2}$ & $-$ &
$\frac{2}{3}$ & 2 & $-$ & $-$ & $-$ & $\frac{1}{6}$ & -$\frac{1}{2}$ \\
\tsp$V_{31}^{N}$ & 2 & $\frac{5}{2}$ & $\frac{43}{3}$ & 11 &
5 & $\frac{40}{3}$ & $\frac{20}{3}$ & $-$ & $-$ & -$\frac{5}{2}$ &
1 & $\frac{13}{3}$ \\
\tsp$V_{31}^{f}$ & 2 & -1 & -$\frac{16}{3}$ & -4 &
-2 & -$\frac{16}{3}$ & -$\frac{8}{3}$ & $-$ & $-$ & 1 & $-$ &
-$\frac{4}{3}$ \\[0.5ex]
\hline\hline
\end{tabular}
\caption[]{Two-loop pole contributions for the $\LR$ four-quark type
diagrams in fig. 10. For $V_{29}$, $V_{30}$ and $V_{31}$,
the results proportional to the number of colour $N$ ($V^{N}$) and
flavour $f$ ($V^{f}$) are separately reported.}
\label{vvlr}
\end{center}
\end{table}

\newpage
\clearpage

\begin{table}
\begin{center}
\begin{tabular}{|c|ccc|ccc|cc|ccc|}
\hline\hline
\multicolumn{1}{|c} { } &
\multicolumn{3}{|c}{$D$} &
\multicolumn{3}{|c}{$C$} &
\multicolumn{2}{|c}{$E$} &
\multicolumn{3}{|c|}{\tsq$\overline{D}$} \\
\hline
\multicolumn{1}{|c|}{$T_N$}&
$\frac{1}{\epdue}$ & $\tsq{\frac{1}{\epuno}}^{\sss HV}$ &
${\frac{1}{\epuno}}^{\sss NDR}$ & $\frac{1}{\epdue}$ &
${\frac{1}{\epuno}}^{\sss HV}$ & ${\frac{1}{\epuno}}^{\sss NDR}$ &
${\frac{1}{\epuno}}^{\sss HV}$ & ${\frac{1}{\epuno}}^{\sss NDR}$ &
$\frac{1}{\epdue}$ & ${\frac{1}{\epuno}}^{\sss HV}$ &
${\frac{1}{\epuno}}^{\sss NDR}$ \\[0.5ex]
\hline
\tsp$P_2$ & $-$ & -$\frac{26}{9}$ & -$\frac{2}{9}$ &
$-$ & $\frac{8}{9}$ & $\frac{32}{9}$ & $-$ & $-$ & $-$ &
$-\frac{34}{9}$ & -$\frac{34}{9}$ \\
\tsp$P_3$ & 4 & $\frac{128}{9}$ & $\frac{92}{9}$ & 8 &
$\frac{112}{9}$ & $\frac{16}{9}$ & $-$ & $-$ & -4 & $\frac{16}{9}$ &
$\frac{76}{9}$ \\
\tsp$P_4$ & -$\frac{23}{9}$ & -$\frac{611}{54}$ & -$\frac{437}{54}$ &
-$\frac{46}{9}$ & -$\frac{308}{27}$ & -$\frac{272}{27}$ & $-$ & $-$ &
$\frac{23}{9}$ & $\frac{5}{54}$ & $\frac{107}{54}$ \\
\tsp$P_{4}^{bg}$ & -$\frac{22}{9}$ & -$\frac{305}{27}$ & -$\frac{221}{27}$ &
-$\frac{44}{9}$ & -$\frac{292}{27}$ & -$\frac{256}{27}$ & $-$ & $-$ &
$\frac{22}{9}$ & -$\frac{13}{27}$ & $\frac{35}{27}$ \\
\tsp$P_{6}$ & $\frac{1}{9}$ & -$\frac{59}{54}$ & -$\frac{65}{54}$ &
$\frac{2}{9}$ & $\frac{52}{27}$ & $\frac{160}{27}$ & $-$ & $-$ &
$-\frac{1}{9}$ & -$\frac{163}{54}$ & -$\frac{385}{54}$ \\
\tsp$P_{6}^{bg}$ & $\frac{8}{3}$ & $\frac{26}{3}$ & 6 & $\frac{16}{3}$ &
$\frac{100}{9}$ & $\frac{112}{9}$ & $-$ & $-$ & -$\frac{8}{3}$ &
-$\frac{22}{9}$ & -$\frac{58}{9}$ \\
\tsp$P_{8}$ & -$\frac{4}{3}$ & -$\frac{22}{3}$ & -6 &
-$\frac{8}{3}$ & -$\frac{16}{9}$ & -$\frac{16}{9}$ & $\frac{4}{3}$
& $-$ & $\frac{4}{3}$ & -$\frac{38}{9}$ & -$\frac{38}{9}$ \\
\tsp$P_{9}$ & $\frac{4}{3}$ & $\frac{16}{3}$ & 4 &
$\frac{8}{3}$ & $\frac{40}{9}$ & $\frac{16}{9}$ & $-$ & $-$ &
-$\frac{4}{3}$ & $\frac{8}{9}$ & $\frac{20}{9}$ \\
\tsp$P_{10}^{LV}$ & -2 & -6 & -4 &
-4 & -$\frac{16}{3}$ & -$\frac{16}{3}$ & $-$ & $-$ & 2 &
-$\frac{2}{3}$ & $\frac{4}{3}$ \\
\tsp$P_{10}^{LA}$ & $-$ & 3 & 3 & $-$ & $-$ & $-$ & $-$ & $-$ & $-$ &
3 & 3 \\
\tsp$P_{11}^{LV}$ & 2 & 6 & 4 &
4 & $\frac{16}{3}$ & $\frac{16}{3}$ & $-$ & $-$ & -2 &
$\frac{2}{3}$ & -$\frac{4}{3}$ \\
\tsp$P_{11}^{LA}$ & $-$ & 3 & 3 & $-$ & $-$ & $-$ & $-$ & $-$ & $-$ &
3 & 3 \\
\tsp$P_{13}$ & $\frac{8}{9}$ & $\frac{139}{27}$ & $\frac{91}{27}$ &
$\frac{16}{9}$ & $\frac{95}{27}$ & $\frac{47}{27}$ & $-$ &
$-$ & -$\frac{8}{9}$ & $\frac{44}{27}$ & $\frac{44}{27}$ \\
\tsp$P_{14}^{LV}$ & 2 & $\frac{27}{3}$ & 9 &
4 & $\frac{22}{3}$ & $\frac{34}{3}$ & $-$ & $-$ & -2 &
$\frac{5}{3}$ & -$\frac{7}{3}$ \\
\tsp$P_{14}^{LA}$ & -2 & -$\frac{31}{3}$ & -7 & -4 & -10 &
-$\frac{22}{3}$ & $-$ & $-$ & 2 & -$\frac{1}{3}$ & $\frac{1}{3}$ \\
\tsp$P_{15}^{LV}$ & -2 & -$\frac{31}{3}$ & -5 &
-4 & -10 & -$\frac{10}{3}$ & $-$ & $-$ & 2 & -$\frac{1}{3}$ & -$\frac{5}{3}$ \\
\tsp$P_{15}^{LA}$ & -2 & -$\frac{31}{3}$ & -7 & -4 & -10 &
-$\frac{22}{3}$ & $-$ & $-$ & 2 & -$\frac{1}{3}$ & $\frac{1}{3}$ \\
\tsp$P_{16}$ & $-$ & $-$ & $-$ &
$-$ & $-$ & $-$ & $\frac{4}{3}$
& $-$ & $-$ & $\frac{4}{3}$ & $-$ \\[0.5ex]
\hline\hline
\end{tabular}
\caption[]{Two-loop pole contributions for the $P$-type penguin diagrams
in fig. 11 when a $\LL$ structure is inserted in the upper vertex.
All penguin diagrams
have a $\LV$ structure, except $P_{10}$, $P_{11}$, $P_{14}$ and $P_{15}$,
for which we explicitly give the $\LV$ ($P^{LV}$) and the $\LA$ ($P^{LA}$)
part.
For $P_{4}$ and $P_{6}$ both the Feynman ($P$) and the background Feynman
($P^{bg}$)
gauge results are reported. Diagrams $P_{5}$, $P_{7}$ and $P_{12}$ are not
included, because their pole parts vanish and they do not contribute to the
two-loop anomalous dimension. Some contributions to
diagram $P_{13}$ are identical for the bare diagram and the counter-term.
This means that one does not need to compute these terms since
they cancel in the final result. For this reason they have not
  been reported in the table.}
\label{pll}
\end{center}
\end{table}

\newpage
\clearpage

\begin{table}
\begin{center}
\begin{tabular}{|c|ccc|ccc|cc|ccc|}
\hline\hline
\multicolumn{1}{|c} { } &
\multicolumn{3}{|c}{$D$} &
\multicolumn{3}{|c}{$C$} &
\multicolumn{2}{|c}{$E$} &
\multicolumn{3}{|c|}{\tsq$\overline{D}$} \\
\hline
\multicolumn{1}{|c|}{$T_N$}&
$\frac{1}{\epdue}$ & $\tsq{\frac{1}{\epuno}}^{\sss HV}$ &
${\frac{1}{\epuno}}^{\sss NDR}$ & $\frac{1}{\epdue}$ &
${\frac{1}{\epuno}}^{\sss HV}$ & ${\frac{1}{\epuno}}^{\sss NDR}$ &
${\frac{1}{\epuno}}^{\sss HV}$ & ${\frac{1}{\epuno}}^{\sss NDR}$ &
$\frac{1}{\epdue}$ & ${\frac{1}{\epuno}}^{\sss HV}$ &
${\frac{1}{\epuno}}^{\sss NDR}$ \\[0.5ex]
\hline
\tsp$F_2$ & $-$ & -$\frac{20}{9}$ & -$\frac{8}{9}$ &
$-$ & $\frac{20}{9}$ & -$\frac{4}{9}$ & $\frac{2}{3}$ & -$\frac{2}{3}$ &
$-$ & -$\frac{34}{9}$ & -$\frac{10}{9}$ \\
\tsp$F_3$ & 4 & $\frac{134}{9}$ & $\frac{122}{9}$ & 8 &
$\frac{124}{9}$ & $\frac{100}{9}$ & $\frac{2}{3}$ & -$\frac{2}{3}$ &
-4 & $\frac{16}{9}$ & $\frac{16}{9}$ \\
\tsp$F_4$ & -$\frac{23}{9}$ & -$\frac{611}{54}$ & -$\frac{575}{54}$ &
-$\frac{46}{9}$ & -$\frac{308}{27}$ & -$\frac{272}{27}$ & $-$ & $-$ &
$\frac{23}{9}$ & $\frac{5}{54}$ & -$\frac{31}{54}$ \\
\tsp$F_{4}^{bg}$ & -$\frac{22}{9}$ & -$\frac{305}{27}$ & -$\frac{287}{27}$ &
-$\frac{44}{9}$ & -$\frac{292}{27}$ & -$\frac{256}{27}$ & $-$ & $-$ &
$\frac{22}{9}$ & -$\frac{13}{27}$ & -$\frac{31}{27}$ \\
\tsp$F_{6}$ & $\frac{1}{9}$ & -$\frac{59}{54}$ & -$\frac{59}{54}$ &
$\frac{2}{9}$ & $\frac{52}{27}$ & $\frac{52}{27}$ & $-$ & $-$ &
$-\frac{1}{9}$ & -$\frac{163}{54}$ & -$\frac{163}{54}$ \\
\tsp$F_{6}^{bg}$ & $\frac{8}{3}$ & $\frac{26}{3}$ & $\frac{26}{3}$ &
$\frac{16}{3}$ & $\frac{100}{9}$ & $\frac{100}{9}$ & $-$ & $-$ &
-$\frac{8}{3}$ & -$\frac{22}{9}$ & -$\frac{22}{9}$ \\
\tsp$F_{8}$ & -$\frac{4}{3}$ & -$\frac{22}{3}$ & -$\frac{22}{3}$ &
-$\frac{8}{3}$ & -$\frac{16}{9}$ & -$\frac{40}{9}$ & $\frac{4}{3}$
& $-$ & $\frac{4}{3}$ & -$\frac{38}{9}$ & -$\frac{26}{9}$ \\
\tsp$F_{9}$ & $\frac{4}{3}$ & $\frac{16}{3}$ & $\frac{16}{3}$ &
$\frac{8}{3}$ & $\frac{40}{9}$ & $\frac{40}{9}$ & $-$ & $-$ &
-$\frac{4}{3}$ & $\frac{8}{9}$ & $\frac{8}{9}$ \\
\tsp$F_{10}^{LV}$ & -2 & -6 & -6 &
-4 & -$\frac{16}{3}$ & -$\frac{16}{3}$ & $-$ & $-$ & 2 &
-$\frac{2}{3}$ & -$\frac{2}{3}$ \\
\tsp$F_{10}^{LA}$ & $-$ & 3 & 3 & $-$ & $-$ & $-$ & $-$ & $-$ & $-$ &
3 & 3 \\
\tsp$F_{11}^{LV}$ & 2 & 6 & 6 &
4 & $\frac{16}{3}$ & $\frac{16}{3}$ & $-$ & $-$ & -2 &
$\frac{2}{3}$ & $\frac{2}{3}$ \\
\tsp$F_{11}^{LA}$ & $-$ & 3 & 3 & $-$ & $-$ & $-$ & $-$ & $-$ & $-$ &
3 & 3 \\
\tsp$F_{13}$ & $\frac{8}{9}$ & $\frac{139}{27}$ & $\frac{115}{27}$ &
$\frac{16}{9}$ & $\frac{95}{27}$ & $\frac{47}{27}$ & $-$ &
$-$ & -$\frac{8}{9}$ & $\frac{44}{27}$ & $\frac{68}{27}$ \\
\tsp$F_{14}^{LV}$ & 2 & $\frac{27}{3}$ & 11 &
4 & $\frac{22}{3}$ & $\frac{34}{3}$ & $-$ & $-$ & -2 &
$\frac{5}{3}$ & -$\frac{1}{3}$ \\
\tsp$F_{14}^{LA}$ & -2 & -$\frac{31}{3}$ & -9 & -4 & -10 &
-$\frac{22}{3}$ & $-$ & $-$ & 2 & -$\frac{1}{3}$ & -$\frac{5}{3}$ \\
\tsp$F_{15}^{LV}$ & -2 & -$\frac{31}{3}$ & -7 &
-4 & -10 & -$\frac{10}{3}$ & $-$ & $-$ & 2 &
-$\frac{1}{3}$ & -$\frac{11}{3}$ \\
\tsp$F_{15}^{LA}$ & -2 & -$\frac{31}{3}$ & -9 & -4 & -10 &
-$\frac{22}{3}$ & $-$ & $-$ & 2 & -$\frac{1}{3}$ & -$\frac{5}{3}$ \\
\tsp$F_{16}$ & $-$ & $-$ & $-$ &
$-$ & $-$ & $-$ & $\frac{4}{3}$
& $-$ & $-$ & $\frac{4}{3}$ & $-$ \\[0.5ex]
\hline\hline
\end{tabular}
\caption[]{Two-loop pole contributions for the $F$-type penguin diagrams
in fig. 11 when a $\LL$ structure is inserted in the upper vertex.
All penguin diagrams
have a $\LV$ structure, except $F_{10}$, $F_{11}$, $F_{14}$ and $F_{15}$,
for which we explicitly give the $\LV$ ($F^{LV}$) and the $\LA$ ($F^{LA}$)
part.
For $F_{4}$ and $F_{6}$ both the Feynman ($F$) and the background Feynman
($F^{bg}$)
gauge results
are reported. Diagrams $F_{5}$, $F_{7}$ and $F_{12}$ are not included,
because their pole parts vanish and they do not contribute to the two-loop
anomalous dimension. Some contributions to
diagram $F_{13}$ are identical for the bare diagram and the counter-term.
This means that one does not need to compute these terms since
they cancel in the final result. For this reason they have not
  been reported in the table.}

\label{fll}
\end{center}
\end{table}

\newpage
\clearpage

\begin{table}
\begin{center}
\begin{tabular}{|c|ccc|ccc|cc|ccc|}
\hline\hline
\multicolumn{1}{|c} { } &
\multicolumn{3}{|c}{$D$} &
\multicolumn{3}{|c}{$C$} &
\multicolumn{2}{|c}{$E$} &
\multicolumn{3}{|c|}{\tsq$\overline{D}$} \\
\hline
\multicolumn{1}{|c|}{$T_N$}&
$\frac{1}{\epdue}$ & $\tsq{\frac{1}{\epuno}}^{\sss HV}$ &
${\frac{1}{\epuno}}^{\sss NDR}$ & $\frac{1}{\epdue}$ &
${\frac{1}{\epuno}}^{\sss HV}$ & ${\frac{1}{\epuno}}^{\sss NDR}$ &
${\frac{1}{\epuno}}^{\sss HV}$ & ${\frac{1}{\epuno}}^{\sss NDR}$ &
$\frac{1}{\epdue}$ & ${\frac{1}{\epuno}}^{\sss HV}$ &
${\frac{1}{\epuno}}^{\sss NDR}$ \\[0.5ex]
\hline
\tsp$F_2$ & -4 & -$\frac{134}{9}$ & -$\frac{122}{9}$ &
-8 & -$\frac{100}{9}$ & -$\frac{172}{9}$ & $\frac{2}{3}$ & -$\frac{10}{3}$ &
4 & -$\frac{28}{9}$ & $\frac{20}{9}$ \\
\tsp$F_3$ & $-$ & $\frac{20}{9}$ & $\frac{8}{9}$ & $-$ &
$\frac{4}{9}$ & -$\frac{68}{9}$ & $\frac{2}{3}$ & -$\frac{10}{3}$ &
$-$ & $\frac{22}{9}$ & $\frac{46}{9}$ \\
\tsp$F_4$ & -$\frac{23}{9}$ & -$\frac{611}{54}$ & -$\frac{575}{54}$ &
-$\frac{46}{9}$ & -$\frac{308}{27}$ & -$\frac{272}{27}$ & $-$ & $-$ &
$\frac{23}{9}$ & $\frac{5}{54}$ & -$\frac{31}{54}$ \\
\tsp$F_{4}^{bg}$ & -$\frac{22}{9}$ & -$\frac{305}{27}$ & -$\frac{287}{27}$ &
-$\frac{44}{9}$ & -$\frac{292}{27}$ & -$\frac{256}{27}$ & $-$ & $-$ &
$\frac{22}{9}$ & -$\frac{13}{27}$ & -$\frac{31}{27}$ \\
\tsp$F_{6}$ & $\frac{1}{9}$ & -$\frac{59}{54}$ & -$\frac{59}{54}$ &
$\frac{2}{9}$ & $\frac{52}{27}$ & $\frac{52}{27}$ & $-$ & $-$ &
$-\frac{1}{9}$ & -$\frac{163}{54}$ & -$\frac{163}{54}$ \\
\tsp$F_{6}^{bg}$ & $\frac{8}{3}$ & $\frac{26}{3}$ & $\frac{26}{3}$ &
$\frac{16}{3}$ & $\frac{100}{9}$ & $\frac{100}{9}$ & $-$ & $-$ &
-$\frac{8}{3}$ & -$\frac{22}{9}$ & -$\frac{22}{9}$ \\
\tsp$F_{8}$ & -$\frac{4}{3}$ & -$\frac{22}{3}$ & -$\frac{22}{3}$ &
-$\frac{8}{3}$ & -$\frac{16}{9}$ & -$\frac{40}{9}$ & $\frac{4}{3}$
& $-$ & $\frac{4}{3}$ & -$\frac{38}{9}$ & -$\frac{26}{9}$ \\
\tsp$F_{9}$ & $\frac{4}{3}$ & $\frac{16}{3}$ & $\frac{16}{3}$ &
$\frac{8}{3}$ & $\frac{40}{9}$ & $\frac{40}{9}$ & $-$ & $-$ &
-$\frac{4}{3}$ & $\frac{8}{9}$ & $\frac{8}{9}$ \\
\tsp$F_{10}^{LV}$ & -2 & -6 & -6 &
-4 & -$\frac{16}{3}$ & -$\frac{16}{3}$ & $-$ & $-$ & 2 &
-$\frac{2}{3}$ & -$\frac{2}{3}$ \\
\tsp$F_{10}^{LA}$ & $-$ & -3 & -3 & $-$ & $-$ & $-$ & $-$ & $-$ & $-$ &
-3 & -3 \\
\tsp$F_{11}^{LV}$ & 2 & 6 & 6 &
4 & $\frac{16}{3}$ & $\frac{16}{3}$ & $-$ & $-$ & -2 &
$\frac{2}{3}$ & $\frac{2}{3}$ \\
\tsp$F_{11}^{LA}$ & $-$ & -3 & -3 & $-$ & $-$ & $-$ & $-$ & $-$ & $-$ &
-3 & -3 \\
\tsp$F_{13}$ & $\frac{8}{9}$ & $\frac{139}{27}$ & $\frac{115}{27}$ &
$\frac{16}{9}$ & $\frac{95}{27}$ & $\frac{47}{27}$ & $-$ &
$-$ & -$\frac{8}{9}$ & $\frac{44}{27}$ & $\frac{68}{27}$ \\
\tsp$F_{14}^{LV}$ & 2 & $\frac{27}{3}$ & 11 &
4 & $\frac{22}{3}$ & $\frac{34}{3}$ & $-$ & $-$ & -2 &
$\frac{5}{3}$ & -$\frac{1}{3}$ \\
\tsp$F_{14}^{LA}$ & -2 & -$\frac{31}{3}$ & -9 & -4 & -10 &
-$\frac{22}{3}$ & $-$ & $-$ & 2 & -$\frac{1}{3}$ & -$\frac{5}{3}$ \\
\tsp$F_{15}^{LV}$ & -2 & -$\frac{31}{3}$ & -7 &
-4 & -10 & -$\frac{10}{3}$ & $-$ & $-$ & 2 &
-$\frac{1}{3}$ & -$\frac{11}{3}$ \\
\tsp$F_{15}^{LA}$ & -2 & -$\frac{31}{3}$ & -9 & -4 & -10 &
-$\frac{22}{3}$ & $-$ & $-$ & 2 & -$\frac{1}{3}$ & -$\frac{5}{3}$ \\
\tsp$F_{16}$ & $-$ & $-$ & $-$ &
$-$ & $-$ & $-$ & $\frac{4}{3}$ & $-$ & $-$ & $\frac{4}{3}$ & $-$ \\[0.5ex]
\hline\hline
\end{tabular}
\caption[]{Two-loop pole contributions for the $F$-type penguin diagrams
in fig. 11 when a $\LR$ structure is inserted in the upper vertex.
All penguin diagrams
have a $\LV$ structure, except $F_{10}$, $F_{11}$, $F_{14}$ and $F_{15}$,
for which we explicitly give the $\LV$ ($F^{LV}$) and the $\LA$ ($F^{LA}$)
part.
For $F_{4}$ and $F_{6}$ both the Feynman ($F$) and the background Feynman
($F^{bg}$)
gauge results
are reported. Diagrams $F_{5}$, $F_{7}$ and $F_{12}$ are not included,
because their pole parts vanish and they do not contribute to the two-loop
anomalous dimension. Some contributions to
diagram $F_{13}$ are identical for the bare diagram and the counter-term.
This means that one does not need to compute these terms since
they cancel in the final result. For this reason they have not
been reported in the table.}
\label{flr}
\end{center}
\end{table}

\end{document}